\documentclass[journal,10 pt,twocolumn]{IEEEtranTCOM}
\usepackage{epsfig}
\usepackage{graphicx}
\usepackage{balance}
\usepackage{float}
\usepackage{psfrag}
\usepackage{amsfonts,amsmath,amssymb,mathrsfs,txfonts}

\newtheorem{lemma}{Lemma}

\usepackage{url}
\usepackage{cite}
\usepackage{verbatim}
\usepackage{multirow}
\usepackage{pifont}
\usepackage{algorithm,algpseudocode}
\usepackage{color}
\usepackage{lettrine}
\usepackage{fancyhdr}
 \fancypagestyle{firststyle}{
 \fancyhf{} 
 \fancyfoot[LO,CE]{\small{2332-7731~\copyright~2025 IEEE. All rights reserved, including rights for text and data mining, and training of artificial intelligence and similar technologies. Personal use is permitted, but republication/redistribution requires IEEE permission. }}
 } 
\begin{document}
\markboth{IEEE Transactions on Cognitive Communications and Networking, DOI: 10.1109/TCCN.2025.3576814 }
{Shell \MakeLowercase{\textit{et al.}}: Bare Demo of IEEEtran.cls for IEEE Journals}

\title{DFRC Systems Co-existing in Licensed Spectrum: Cognitive Beamforming Designs}
\author{Tuan Anh Le, \IEEEmembership{Senior, IEEE},
Ivan Ku, Xin-She Yang, Christos Masouros, \IEEEmembership{Fellow, IEEE} and Tho Le-Ngoc,
\IEEEmembership{Life Fellow, IEEE}
\thanks{An earlier version of this paper was presented in part at the IEEE Vehicular Technology
Conference (VTC 2024-Spring), Singapore, June 24-27, 2024 [DOI: 10.1109/VTC2024-Spring62846.2024.10683337].}
\thanks{T. A. Le and X.-S. Yang are with the Faculty of Science and Technology, Middlesex University, London, NW4 4BT, UK. (email: \{t.le; x.yang\}@mdx.ac.uk)}
\thanks{I. Ku is with the Faculty of Artificial Intelligence and Engineering, Multimedia University, Persiaran Multimedia
63100, Cyberjaya, Selangor, Malaysia (e-mail: kccivan@mmu.edu.my)}
\thanks{C. Masouros is with the Department of Electronic and Electrical Engineering, University College London, London, WC1E 7JE, UK. (email: c.masouros@ucl.ac.uk)}
\thanks{T. Le-Ngoc is with the Department of Electrical and Computer Engineering, McGill University, Montreal, Quebec, H3A 0G4, Canada  (e-mail: tho.le-ngoc@mcgill.ca)}}
\maketitle
\thispagestyle{firststyle}
\begin{abstract}
This paper introduces a dual-function radar-communication (DFRC) system with cognitive radio capability to tackle the spectral scarcity problem in wireless communications. Particularly, a cognitive DFRC system operates on a spectrum owned by a primary system to simultaneously perform data communication and target tracking with the condition that its interference to the primary users (PUs) is below a certain threshold. To achieve this, an optimization problem is formulated to jointly design the beamforming vectors for both the radar and communication functions in such a way that the mean square error (MSE) of the beam pattern between the designed and desired waveforms is minimized. The optimization problem has the following three constraints: i) the signal-to-interference-plus-noise ratio (SINR) at each data communication user is above a predetermined level; ii) the per-antenna transmit power is maintained at a given level; iii) the interference imposed on each PU is below a certain threshold. Both the semidefinite relaxation and nature-inspired firefly algorithms are proposed in order to search for the optimal solutions to the optimization problem. The simulation results indicate that our proposed algorithms can enable the DFRC system to protect the PUs while simultaneously performing its communication and radar functions.
\end{abstract}



\section{Introduction}
\label{sec:introduction}
Increasing demand of bandwidth in wireless networks has been the driving force behind the development of spectrum sharing technologies. Cognitive radio was introduced and developed for effective spectrum sharing and efficient spectrum utilization in band-limited wireless communications systems. In cognitive radio, access to spectrum owned by a primary system is granted to the secondary system if the latter's operation either imposes acceptable interference level on the former's primary users (PUs) or does not affect the PUs at all, see e.g., \cite{Keivan2010,TuanTcom2014} and references therein. Furthermore, the dual-function radar-communications (DFRC) system has recently attracted increasing interests as a means of sharing spectrum, hardware, and signalling between radar and communication systems. Consequently, DFRC has been recognized as a promising technique in beyond 5G and 6G networks \cite{Zhang10005142,Liu8386661,Xiang2020,Liu9737357}.

DFRC dates back to 1963 when the idea of using radar pulse codes to convey information on top of the tracking signal was developed \cite{1stDFRC}. The phase \cite{HassanienIET} and amplitude \cite{Hassanien7347464} of the radar's spatial side-lobes were utilized to embed information bits, while the authors in \cite{Wang8438940} employed antenna selection techniques to achieve this. The methods above are regarded as radar-centric DFRC where the radar function takes precedence and information bits are enclosed within some parameters of the radar waveform. On the other hand, communication-centric DFRC emphasizes the communication function over the radar function \cite{Zhang10005142}. In \cite{Liu8386661}, dual-functional waveforms were jointly designed where flexible tradeoffs between the communication and radar functions are obtained. Besides that, communication waveforms for orthogonal frequency division multiplexing and direct sequence spread spectrum were exploited in \cite{Sturm5776640} for the dual role of target detection.

In order to balance the radar's and communication's performances in the DFRC system, the radar function and communication function can utilize separate waveforms that are jointly designed \cite{Xiang2020}, also known as the beamforming design approach. In this approach, one of the common metrics for the radar function is the mean square error (MSE) of the beam pattern between the designed waveform and the desired waveforms \cite{Xiang2020,Ren10153696}. Alternatively, it can be the minimum weighted beam pattern gain in the desired directions of the targets \cite{Hua10086626} or the cross correlation of the beam pattern \cite{Xiang2020}. Conversely, the metrics for the communication function are commonly the signal-to-interference-plus-noise ratio (SINR) \cite{Xiang2020,Hua10086626,Liu9724205}. The beamforming design problems for DFRC are then cast as optimization problems. For example, the weighted sum of the MSE beam pattern and the cross correlation beam pattern is minimized subject to the communication users' SINR and transmit power constraints as proposed in \cite{Xiang2020}, while, in \cite{Hua10086626}, the MSE beam pattern is minimized under the constraints of the communication users' SINR and the transmit power. In \cite{Liu9724205}, the objective is to maximize the worst SINR among the users subject to the power constraint and the covariance of the transmit waveform being equal to a given optimal covariance of the multiple-input multiple-output (MIMO) radar. 

Non-convex beamforming problems in DFRC systems are normally transformed into quadratic semidefinite programmings (QSDPs) by employing semidefinite relaxation (SDR) technique, see e.g., \cite{Liu8386661,Xiang2020,Ren10153696,Hua10086626}. In SDR technique, a new optimization variable  is introduced in a form of a positive semidefinite matrix which is the product of an original optimization vector and its conjugate transpose, see e.g., \cite{Zhi,Huang2010new,Tuan10214071} and references therein. As the newly introduced optimization matrix has only one linearly independent column, an additional non-convex constraint of rank one is imposed on the QSDP. Relaxing the rank-one constraint results in a convex QSDP which can be effectively solved by interior-point methods \cite{Boyd_convex}. The rank-one relaxation problem is tight\footnote{The optimum solution to the rank-one relaxation QSDP is also the optimal solution to the original rank-one constraint QSDP.} if solving the rank-one relaxation QSDP yields a rank-one optimal solution. Otherwise, only sub-optimal solution to the original problem can be attained by utilizing a randomization technique \cite{Wei11}.

Nature-inspired algorithms, also known as meta-heuristic algorithms, have been developed based on some abstractions of nature to find solutions for optimization problems. Meta-heuristic algorithms possess two important characteristics, i.e., exploitation and exploration. The exploitation characteristic allows a nature-inspired algorithm to converge to the optimal solution while the exploration characteristic helps the algorithm to escape from being trapped at local optimums and also to increase the diversity of the solutions \cite{YangFA2008,YangFA2009}. Owning the exploitation and exploration characteristics, Firefly algorithm (FA), which was developed in late 2007 and introduced in 2008 by Yang \cite{YangFA2008,YangFA2009}, can effectively solve non-convex multimodal problems. The original FA was designed for optimization problems with single and scalar optimization variable. Recently, the original FA has been generalized into a framework to capture multivariate optimization problems where optimization variable can be either scalars, vectors, matrices or the combination of all \cite{Tuan10311527}. The generalized FA framework outperforms the interior-point method for various transmit beamforming problems in contemporary wireless communications systems such as reconfigurable intelligent surfaces \cite{Tuan10201127} and movable antenna arrays \cite{TuanWCL2024}. Thanks to the generalized nature of the FA framework in \cite{Tuan10311527}, its applications are beyond transmit beamforming problems.

The combination of DFRC with cognitive radio offers additional improvement to the spectrum utilization. Cognitive radio was firstly adopted in DFRC in \cite{Rathapon6331681} where opportunistic spectrum sharing between a primary rotating radar system and a secondary communication system was allowed. In fact, this approach can be considered as a radar-centric method where the performance of the communication function can be potentially compromised. Hence, to improve the performance of the communication function while maintaining the performance of the radar function in a cognitive DFRC system, the beamforming design approach is desirable. To that end, we recently proposed a joint design beamforming approach for a cognitive DFRC system in \cite{Tuan_VTC2024Spring} using the SDR technique. This paper further develops our work in \cite{Tuan_VTC2024Spring} by introducing a nature-inspired approach for the cognitive DFRC system. Furthermore, extensive simulation, in comparison with \cite{Tuan_VTC2024Spring}, have been carried out to evaluate the proposed approaches. The contribution of this work can be summarized as follows:
\begin{itemize}
    \item An optimization problem for a cognitive DFRC system is proposed whereby the MSE of the beam pattern between the designed waveforms and the desired waveforms is minimized subject to i) the SINR level of each secondary user at above a required level; ii) the per-antenna transmit power at a fixed level; iii) the interference level imposed on each primary user at below a predefined threshold.
    \item Two approaches, i.e., SDR and nature-inspired FA, are proposed to find the optimal solution to the optimization problem. The FA approach proposed in this work is the generalization of the FA framework in \cite{Tuan10311527} into a novel cognitive DFRC problem.
    \item Complexity analysis is then provided for the proposed SDR and FA approaches.
\end{itemize}
 The problem introduced in this paper differs from that in \cite{Xiang2020} due to the interference constraint imposed to protect the primary users. Furthermore, the FA approach introduced here extends the contribution of the previous work in \cite{Xiang2020}.
 
\emph{\textbf{Notation}:} 
Lower or upper case letter $a$ or $A$: a scalar; bold lower case letter $\mathbf{a}$: a column vector; bold upper case letter $\mathbf{A}$: a matrix; $(\cdot)^T$: the transpose operator; $(\cdot)^H$: the complex conjugate transpose operator; $\left\|\cdot\right\|$: the Euclidean norm; $\mathbb{E}[\cdot]$: the expected value operator; $\textrm{Tr}\left(\cdot\right)$: the trace operator; $\mathbf{A}\succeq \mathbf{0}$: $\mathbf{A}$ is positive semidefinite; $\mathbf{I}_x$: an $x \times x$ identity matrix;   $\mathcal{O}$: the big O notation; $\mathbb{H}^{M\times M}$: the set of $M\times M$ Hermitian matrices;  $\mathbb{C}^{M\times 1}$: the set of $M\times 1$ complex element vectors; $a\sim\mathcal{CN}(0,\sigma^2)$: $a$ is the zero mean circularly symmetric complex Gaussian random variable with variance $\sigma^2$.
\section{System Model}
\begin{table*}[ht]
  \centering
  	\begin{tabular}{|c|l|c|}
            \hline
            \textbf{Symbol/Term}  & \textbf{Description}  & \textbf{Dimension}\\
            \hline
            $U$ & Number of single-antenna secondary users (SUs) & $\mathbb{Z}^{1 \times 1}$\\
            \hline
            $K$ & Number of simultaneously tracked targets & $\mathbb{Z}^{1 \times 1}$\\
            \hline
            $L$ & Number of primary users (PUs) & $\mathbb{Z}^{1 \times 1}$\\
            \hline
            $M$ & Number of antennas at the MIMO base station & $\mathbb{Z}^{1 \times 1}$\\
            \hline
            $\mathbf{w}_i$ & Transmit beamforming vector for the $i$-th SU & $\mathbb{C}^{M \times 1}$\\
            \hline
            $x_i^c$ & Intended data symbol for the $i$-th SU & $\mathbb{C}^{1 \times 1}$\\
            \hline
            $\mathbf{v}_t$ & Radar beamforming vector for the $t$-th radar target & $\mathbb{C}^{M \times 1}$\\
            \hline
            $x_t^r$ & Radar waveform for the $t$-th radar target & $\mathbb{C}^{1 \times 1}$\\
            \hline
            $\mathbf{x}$ & Signal output vector of the $M$-antenna array & $\mathbb{C}^{M \times 1}$\\
            \hline
            $p_i$ & Transmit power allocated to the $i$-th SU& $\mathbb{R}^{1 \times 1}$\\
            \hline
            $\mathbf{R}$ & Covariance matrix of the signal output vector & $\mathbb{C}^{M \times M}$\\
            \hline
            $\theta$ & Direction of the beamforming pattern of the signal output, $\mathbf{x}$ & $\mathbb{R}^{1 \times 1}$\\
            \hline
            $\mathcal{P}\left(\mathbf{R},\theta \right)$ & Beamforming pattern of the signal output, $\mathbf{x}$ & $\mathbb{R}^{1 \times 1}$\\
            \hline
            $\mathcal{L}\left(\mathbf{R},\omega \right)$ & MSE between the designed and desired radar beam patterns for all targets & $\mathbb{R}^{1 \times 1}$\\
            \hline
            $\mathcal{D}\left(\theta_{g,t} \right)$ & Desired radar beam pattern for the $t$-th target at the $g$-th angle sample & $\mathbb{R}^{1 \times 1}$\\
            \hline
            $G$ & Beam grid size & $\mathbb{Z}^{+ 1 \times 1}$\\
            \hline
            $\omega$& Optimizable scaling factor for $\mathcal{D}\left(\theta_{g,t} \right)$ & $\mathbb{R}^{1 \times 1}$\\
            \hline
            $y_i$ & Received signal at the $i$-th SU & $\mathbb{C}^{1 \times 1}$\\
            \hline
            $\mathbf{h}_{s,i}$ & Channel coefficient between the BS and the $i$-th SU & $\mathbb{C}^{M \times 1}$\\
            \hline
            $n_i$ & Complex additive white Gaussian noise at the $i$-th SU & $\mathbb{C}^{1 \times 1}$\\
            \hline
            $\mathbf{h}_{p,l}$ & Channel coefficient between the BS and the $l$-th PU & $\mathbb{C}^{M \times 1}$\\
            \hline
            $I_l$ & Interference imposed by the DFRC transmission on the $l$-th PU & $\mathbb{R}^{+ 1 \times 1}$\\
            \hline
            SINR$_i$ & The SINR at the $i$-th SU & $\mathbb{R}^{+ 1 \times 1}$\\
            \hline
            $P_m$ & Available transmit power at the BS & $\mathbb{R}^{+ 1 \times 1}$\\
            \hline
            $I_\text{thres}$ & Interference tolerance threshold at PUs & $\mathbb{R}^{+ 1 \times 1}$\\
            \hline
            $\eta_i$ & Required SINR level of the $i$-th SU & $\mathbb{R}^{+ 1 \times 1}$\\
            \hline
            $\lambda_i$ & Penalty constant for the $i$-th SINR constraint & $\mathbb{R}^{+ 1 \times 1}$\\
            \hline
            $\rho_i$ & Penalty constant for the $i$-th power constraint & $\mathbb{R}^{+ 1 \times 1}$\\
            \hline
            $\zeta_l$ & Penalty constant for the $l$-th interference constraint & $\mathbb{R}^{+ 1 \times 1}$\\
            \hline
            $\beta_0$ & The attractiveness at zero distance between any two fireflies & $\mathbb{R}^{+ 1 \times 1}$\\
            \hline
            $\gamma$ & Attractiveness variation, a.k.a., light
            absorption coefficient & $\mathbb{R}^{+ 1 \times 1}$\\
            \hline
            $\alpha$ & Randomization factor & $\mathbb{R}^{+ 1 \times 1} \in [0,1]$\\
            \hline
            $\Theta$ & Number of FA iteration & $\mathbb{Z}^{ +1 \times 1}$\\
            \hline
            $N$ & Firefly population & $\mathbb{Z}^{+ 1 \times 1}$\\
            \hline
        \end{tabular}
  \caption{Frequently used symbols and terms.}
  \label{tab:2}
\end{table*}
\begin{figure}[t]
\centering
    \includegraphics[width=.45\textwidth]{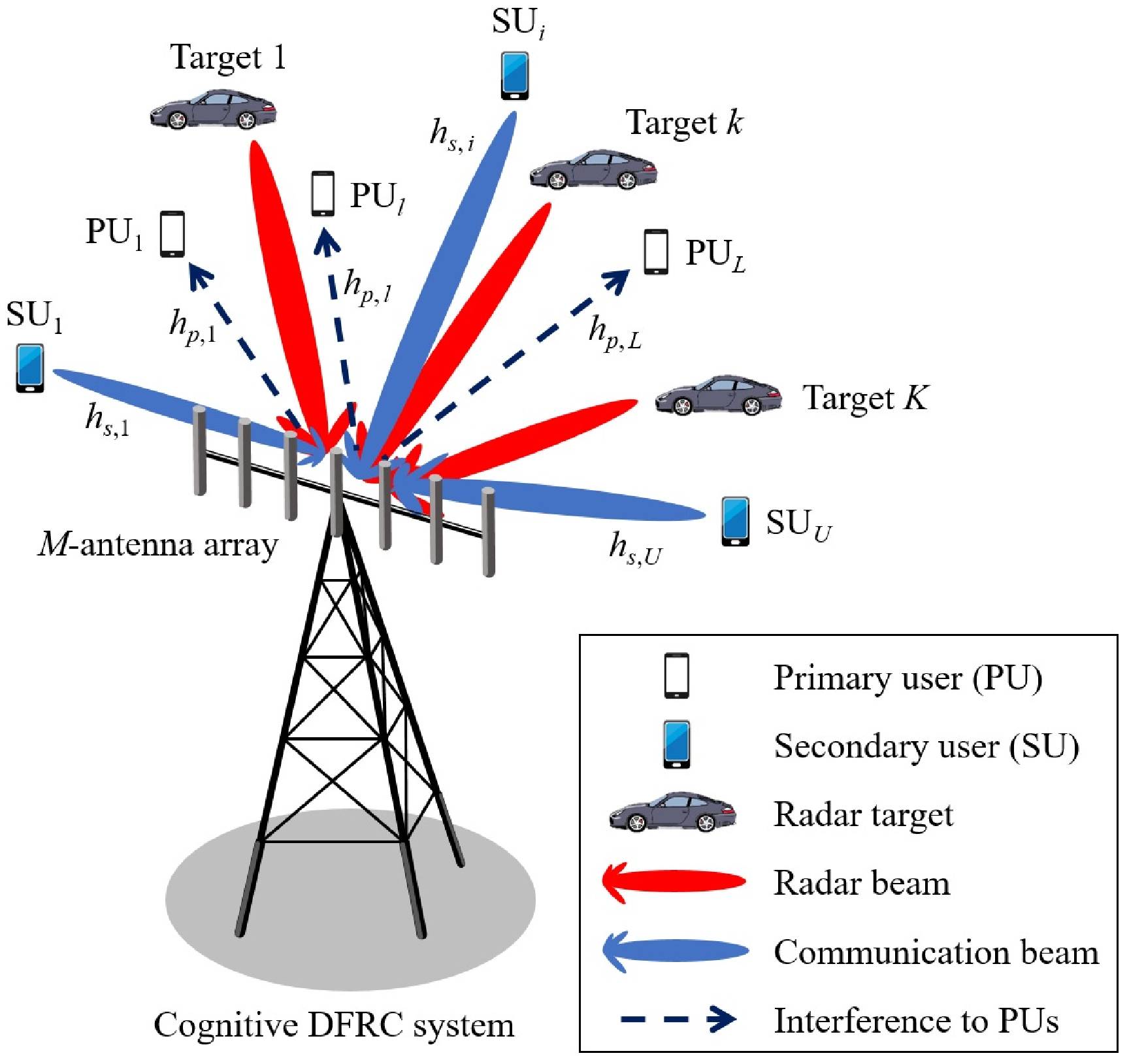}
\caption{A cognitive dual-function radar-communications (DFRC) system.}\label{DFRCSysMdl}
\end{figure}
Consider a DFRC system consisting of an integrated MIMO base station (BS) and the MIMO radar subsystem for data communication and target tracking, respectively, as shown in Fig.~\ref{DFRCSysMdl}. An array of $M$ antennas is shared between the MIMO BS and the MIMO radar subsystem. The DFRC system does not own any radio spectrum, but operates cognitively on the spectrum owned by the primary system. Under primary system permission, the DFRC system communicates with $U$ single-antenna secondary users (SUs) while simultaneously tracking $K$ targets, as long as its interference levels imposed on the $L$ primary users (PUs) are kept below predefined thresholds. 

Let $\mathbf{w}_i \in \mathbb{C}^{M \times 1}$ and $x_i^c$ be the transmit beamforming vector and the intended data symbol, respectively, for the communication function of the $i$-th SU, where $\mathbb{E}\begin{bmatrix}|x_i^c|^2\end{bmatrix}=1$ and $i \in \{1, \cdots, U\}$. While, $\mathbf{v}_t \in \mathbb{C}^{M \times 1}$ and $x_t^r$ represent the radar beamforming vector and radar waveform, respectively, for the radar function of tracking targets, where $\mathbb{E}\begin{bmatrix}|x_t^r|^2\end{bmatrix}=1$ and $t \in \{1, \cdots, M\}$, that is, we have $M$ radar beams tracking $K$ targets. It is assumed that the resource and user scheduling approaches, see e.g., \cite{Memisoglu,Cao2023,Abanto-Leon} and references therein, are exploited to maintain the condition $U+K+L \leq M$ so that there is enough spatial degree of freedom to accommodate the beamforming requirements of all PUs, SUs, and targets that are under consideration. Furthermore, let $\mathbf{W}=\begin{bmatrix}\mathbf{w}_1, \mathbf{w}_2,\cdots,\mathbf{w}_U  \end{bmatrix}$, $\mathbf{x}^c=\begin{bmatrix} {x}^c_1, {x}^c_2,\cdots,{x}^c_U  \end{bmatrix}^T$, $\mathbf{V}=\begin{bmatrix} \mathbf{v}_1, \mathbf{v}_2, \cdots, \mathbf{v}_M  \end{bmatrix}$ and $\mathbf{x}^r=\begin{bmatrix}{x}^r_1, {x}^r_2, \cdots, {x}^r_M  \end{bmatrix}^T$. Following that, the $M \times 1$ signal output vector of the $M$-antenna array can, thus, be written as:
\begin{equation}
    \mathbf{x}=\mathbf{W}\mathbf{x}^c+\mathbf{V}\mathbf{x}^r.
\end{equation}

Finally, the transmit power, $p_i$, allocated to the $i$-th SU is given by:
\begin{equation}
    p_i=\begin{bmatrix}
        \sum_{i=1}^U\mathbf{w}_i\mathbf{w}_i^H+\sum_{t=1}^M\mathbf{v}_t \mathbf{v}_t^H
    \end{bmatrix}_{i,i}=\begin{bmatrix}
        \mathbf{W}\mathbf{W}^H+\mathbf{V} \mathbf{V}^H
    \end{bmatrix}_{i,i},
\end{equation}
where $\begin{bmatrix}
    \mathbf{A}
\end{bmatrix}_{i,i}$ is the $i$-th entry on the diagonal of the matrix $\mathbf{A}$.

\subsection{Radar Metric}
Assuming the communication data symbols and radar waveforms are uncorrelated, that is, $\mathbb{E}\begin{bmatrix}x_i^c x_t^r\end{bmatrix}=0$, we define the covariance matrix of the output vector as
\begin{eqnarray}
    \mathbf{R}&=&\mathbb{E}\left( \mathbf{x}\mathbf{x}^H\right)=\mathbf{W}\mathbf{W}^H+\mathbf{V}\mathbf{V}^H, \nonumber\\
    &=& \sum_{i=1}^{U}\mathbf{w}_i\mathbf{w}_i^H+\sum_{t=1}^{M}\mathbf{v}_t\mathbf{v}_t^H.\label{CoVa}
\end{eqnarray}
The major task of a radar function is to steer its radar beams towards predefined directions so that the signal bounced back from the targets can be analyzed \cite{Xiang2020}. The beamforming pattern of the signal output from the DFRC system at the direction $\theta\in \left[-180^{\circ},180^{\circ} \right]$ can be expressed as:
\begin{eqnarray}
\mathcal{P}\left(\mathbf{R},\theta \right)=\mathbb{E}\left[\Bigg|\mathbf{a}^H\left( \theta\right)\mathbf{x} \Bigg|^2 \right]=\mathbf{a}^H\left( \theta\right) \mathbf{R}\mathbf{a}\left( \theta\right),\label{Beam_patt}
\end{eqnarray}
where the steering vector at angle $\theta$ is $\mathbf{a}^H\left( \theta\right)=\begin{bmatrix}
    1, e^{j2\pi\frac{d}{\lambda}\sin \theta}, & \cdots , & e^{j2\pi\frac{d(M-1)}{\lambda}\sin \theta}
\end{bmatrix}$, the antenna spacing is $d$ and the carrier wavelength is $\lambda$.

One of the important criteria in radar beamforming is to match the designed and desired radar beam patterns \cite{Xiang2020,Stoica}. The MSE between the designed and desired radar beam patterns for all targets is the performance metric of the radar function in the DFRC system \cite{Xiang2020,Stoica} and is given by:
\begin{eqnarray}
    \mathcal{L}\left( \mathbf{R},\omega\right)=\frac{1}{G}\sum_{g=1}^G\sum_{t=1}^K\Bigg |\omega \mathcal{D}\left(\theta_{g,t} \right)-\mathcal{P}\left(\mathbf{R},\theta_{g,t} \right)\Bigg |^2, \label{MSE}
\end{eqnarray}
where $\mathcal{D}\left(\theta_{g,t} \right)$ is the desired radar beam pattern for the $t$-th target at the $g$-th angle sample, $\{ \theta_{g,t}\}_{g=1}^G$ are the sampled angle grids for the $t$-th target and $\omega$ is the scaling factor which is a variable to be optimized. The role of $\omega$ is to properly scale $\mathcal{D}\left(\theta_{g,t} \right)$ as the desired radar beam patterns are usually given in normalized forms \cite{Stoica}. The designed and desired radar beam patterns are said to be matched when the MSE defined in \eqref{MSE} is minimized.

\subsection{Communication Metric}
The received signal at the $i$-th SU is:
\begin{eqnarray}\label{re01}
y_i=\mathbf{h}_{s,i}^H \mathbf{x}+n_i =\mathbf{h}_{s,i}^H \left( \sum_{i=1}^U\mathbf{w}_i {x}_i^c+\sum_{t=1}^M\mathbf{v}_t {x}_t^r \right)+ n_i,
\end{eqnarray}
where $\mathbf{h}_{s,i} \in \mathbb{C}^{M \times 1}$ denotes the $M$ channel coefficients between the BS and the $i$-th SU and  $n_i\sim \mathcal{CN}(0,\sigma^2)$ is the additive noise at the $i$-th SU. Let $\{\mathbf{w}_{i}\}=\{\mathbf{w}_1, \mathbf{w}_2,\cdots, \mathbf{w}_U\}$ and $\{\mathbf{v}_{t}\}=\{\mathbf{v}_1, \mathbf{v}_2,\cdots, \mathbf{v}_M\}$ represent two sets of beamforming vectors. In addition to that, let $\mathbf{H}_{s,i}=\mathbf{h}_{s,i}\mathbf{h}_{s,i}^H$ be the instantaneous channel state information (CSI) or $\mathbf{H}_{s,i}=\mathbb{E}[\mathbf{h}_{s,i}\mathbf{h}_{s,i}^H]$ be the statistical CSI. We assume that the BS has CSI for downlink beamforming. In practice, it is possible to obtain downlink CSI efficiently by exploiting the channel reciprocity property in a time-division-duplex system, see, e.g., \cite{Xiang2020,Yang10605608,Tuan_CL2018,Tuan_TVT_2020,Zhu_TDD_2025,Yang_TDD_2025,Zhang_TDD_2024}, or via dedicated uplink-pilot channels in a frequency-division-duplex system, see, e.g., \cite{ZhaoFDD,GAOFDD,Wesemann_FDD_2017,Xiang_FDD_2023}. Imperfect CSI is beyond the scope of this work and deserves a separate study. The SINR at the $i$-th SU can therefore be written as:
\begin{equation}\label{sinr01}
\textrm{SINR}_{i}=\frac{\mathbf{w}_i^H\mathbf{H}_{s,i}^H \mathbf{w}_i}{ \sum_{j=1,j\neq i}^{U}\mathbf{w}_j^H\mathbf{H}_{s,i}^H\mathbf{w}_j+\sum_{t=1}^{M}\mathbf{v}_t^H\mathbf{H}_{s,i}^H \mathbf{v}_t+\sigma_i^2}.\nonumber
\end{equation}

Similarly, let $\mathbf{h}_{p,l} \in \mathbb{C}^{M \times 1}$ denote the channel coefficients between the BS and the $l$-th PU and  $\mathbf{H}_{p,l}=\mathbf{h}_{p,l}\mathbf{h}_{p,l}^H$ be the instantaneous CSI or $\mathbf{H}_{p,l}=\mathbb{E}[\mathbf{h}_{p,l}\mathbf{h}_{p,l}^H]$ be the statistical CSI. The interference imposed by the DFRC transmission on the $l$-th PU can, thus, be expressed as:
\begin{eqnarray}
   I_{l}= \sum_{j=1}^{U} \mathbf{w}_j^H\mathbf{H}_{p,l}^H\mathbf{w}_j+\sum_{t=1}^{M}\mathbf{v}_t^H\mathbf{H}_{p,l}^H\mathbf{v}_t.\nonumber
\end{eqnarray}

The SINR is a metric to measure the quality of service that is provided to each SU, while, the interference is another metric to measure how much deterioration the DFRC system is causing to the signal reception quality of the PUs.

\section{Proposed Optimization Problem}
Let $P_m$, $I_{\text{thres}}$ and $\eta_i$ be the available transmit power at the BS, the interference tolerance threshold at PUs and the required SINR level of the $i$-th SU, respectively. We aim to jointly design the two sets of beamforming vectors $\{\mathbf{w}_{i}\}$ and $\{\mathbf{v}_{t}\}$ so that the MSE in \eqref{MSE} is minimized subject to the following three constraints: i) the SINR level at each SU is guaranteed to be above $\eta_i$; ii) the transmit power at each BS antenna does not exceed the BS's budget $\frac{P_m}{M}$; iii) the interference level inflicted at each PU is kept below $I_{\text{thres}}$. To that end, we introduce the optimization problem as follows:
\begin{equation} \label{proposedDFRC}
\begin{aligned}
& \underset{ \{\mathbf{w}_{i} \},\ \{\mathbf{v}_{t} \},\ \omega }{\textrm{min}} & &
\mathcal{L}\left( \{\mathbf{w}_{i}\},\{\mathbf{v}_{t}\},\omega\right)\\
& \mbox{s.\ t.}\ & & \textrm{SINR}_{i}\geq \eta_i, \forall i, \\
&&& \begin{bmatrix}
        \mathbf{W}\mathbf{W}^H+\mathbf{V} \mathbf{V}^H
    \end{bmatrix}_{i,i}\leq \frac{P_m}{M}, \ \forall i, \\
&&& \sum_{j=1}^{U} \mathbf{w}_j^H\mathbf{H}_{p,l}^H\mathbf{w}_j+\sum_{t=1}^{M}\mathbf{v}_t^H\mathbf{H}_{p,l}^H\mathbf{v}_t \leq I_{\text{thres}}, \ \forall l.
\end{aligned}
\end{equation}
It can be observed that the optimization problem shown in \eqref{proposedDFRC} is non-convex due to the SINR constraint. To find the optimal solution to this optimization problem, we first introduce the SDR approach. This is then followed by the FA approach.

\section{Proposed SDR Approach}
By letting $\mathbf{F}_i=\mathbf{w}_i\mathbf{w}_i^H$ and $\mathbf{V}_t=\mathbf{v}_t\mathbf{v}_t^H$, one can rewrite \eqref{CoVa} as:
\begin{equation}
    \mathbf{R}=\sum_{i=1}^{U}\mathbf{F}_i + \sum_{t=1}^{M}\mathbf{V}_t.
\end{equation}
In the sequel, we will cast the objective function in a quadratic form with respect to the optimization variables $\mathbf{R}$ and $\omega$. We start by rewriting \eqref{Beam_patt} as follows:
\begin{eqnarray}
\mathcal{P}\left(\mathbf{R},\theta \right)&=&\mathbf{a}^H\left( \theta\right) \mathbf{R}\mathbf{a}\left( \theta\right) \nonumber\\
&=&\textrm{vec}\left( \mathbf{a}^H\left( \theta\right) \mathbf{R}\mathbf{a}\left( \theta\right)\right) \nonumber\\
&=&\left( \mathbf{a}^T\left( \theta\right) \otimes\mathbf{a}^H\left( \theta\right)\right) \textrm{vec}\left(  \mathbf{R}\right).
\end{eqnarray}
Therefore,
\begin{eqnarray}
    \omega \mathcal{D}\left(\theta_{g,t} \right) - \mathcal{P}\left(\mathbf{R},\theta_{g,t} \right) &=&
    \begin{bmatrix}
        \mathcal{D}\left(\theta_{g,t} \right), \ -  \mathbf{a}^T\left( \theta_{g,t}\right) \otimes\mathbf{a}^H\left( \theta_{g,t}\right)
    \end{bmatrix} \nonumber\\
    &\times& \begin{bmatrix}
        \omega \\ \textrm{vec}\left(  \mathbf{R}\right)
    \end{bmatrix}.
\end{eqnarray}
Hence, one can equivalently cast \eqref{MSE} as:
\begin{eqnarray}
    \mathcal{L}\left( \mathbf{R},\omega\right)=\mathbf{r}^H\mathbf{\Omega}\mathbf{r},
\end{eqnarray}
where
\begin{eqnarray}
   \mathbf{r}=\begin{bmatrix}
    \omega \\ \textrm{vec}\left(  \mathbf{R}\right)
    \end{bmatrix}
    \end{eqnarray}
stacks all optimization variables in a vector and 
\begin{eqnarray}
    \mathbf{\Omega} &=& \frac{1}{G} \sum_{g=1}^G \sum_{t=1}^K \begin{bmatrix}
        \mathcal{D}\left(\theta_{g,t} \right) \\ -  \mathbf{a}^{\star}\left( \theta_{g,t}\right) \otimes\mathbf{a}\left( \theta_{g,t}\right)
    \end{bmatrix} \nonumber\\
    &\times& \begin{bmatrix}
        \mathcal{D}\left(\theta_{g,t} \right), \ -  \mathbf{a}^T\left( \theta_{g,t}\right) \otimes\mathbf{a}^H\left( \theta_{g,t}\right)\label{OmegaMatrix}
    \end{bmatrix}.
\end{eqnarray}
With some manipulations, one can equivalently rewrite \eqref{proposedDFRC} as:
\begin{equation}
\begin{aligned}\label{prob_ISAC}
& \underset{  \mathbf{F}_i, \ \mathbf{V}_t}{\textrm{min}}  & & \mathbf{r}^H\mathbf{\Omega}\mathbf{r}\\
& \mbox{s. \ t.}\ & &\left(1+\frac{1}{\eta_i} \right) \textrm{Tr}\left(\mathbf{H}_{s,i}^H\mathbf{F}_{i}\right)-\sum_{j=1}^U\textrm{Tr}
\left(\mathbf{H}_{s,i}^H\mathbf{F}_{j}\right)\\ &&& -\sum_{t=1}^M\textrm{Tr}
\left(\mathbf{H}_{s,i}^H\mathbf{V}_{t}\right)-\sigma^2_i\geq0, \ \forall i,\\
&&& \mathbf{R}=\sum_{i=1}^{U}\mathbf{F}_i + \sum_{t=1}^{M}\mathbf{V}_t,\\
&&& \mathbf{R}_{i,i}\leq\frac{P_m}{M}, \ \forall i, \\
&&& \sum_{j=1}^U\textrm{Tr}\left( \mathbf{H}_{p,l} \mathbf{F}_{j}  \right)+\sum_{t=1}^M\textrm{Tr}\left( \mathbf{H}_{p,l} \mathbf{V}_{t}  \right) \leq I_{\text{thres}}, \ \forall l,\\
&&& \mathbf{F}_i  \succeq \mathbf{0}, \ \forall i  \in \{1,\cdots,U\},\\
&&& \mathbf{V}_t \succeq \mathbf{0},  \ \forall t  \in \{1,\cdots,M\}.
\end{aligned}
\end{equation}
To arrive at \eqref{prob_ISAC}, we have relaxed rank-one constraints on $\mathbf{F}_i$ and $\mathbf{V}_t$, i.e., $\mathrm{rank}(\mathbf{F}_i) = 1$ and $\mathrm{rank}(\mathbf{V}_t) = 1$. The optimization problem in \eqref{prob_ISAC} is convex. Hence, the interior-point methods can be adopted to solve the problem, e.g., using CVX package \cite{cvx2015}. The computational complexity of solving \eqref{prob_ISAC} is stated in the following lemma.
\begin{lemma} \label{SDR_Com_lemma}
The dominant computation complexity to obtain an optimal solution to problem \eqref{prob_ISAC} is dominated by the order of $GKM^{8.5}$.
\end{lemma}
\begin{proof}
See Appendix~\ref{SDR_proof}.
\end{proof}

If the optimal solutions obtained by solving \eqref{prob_ISAC} are rank-one matrices $\mathbf{F}_i$ and $\mathbf{V}_t$, then they are also the optimal solutions to the original optimization problem in \eqref{proposedDFRC} where the optimal solutions $\mathbf{w}_i$ and $\mathbf{v}_t$ are attained from the product of the square root of the eigenvalue and the eigenvector of the corresponding $\mathbf{F}_i$ and $\mathbf{V}_t$ \cite{Tuan10214071,TuanTcom2013}. On the other hand, if the optimal solution $\mathbf{F}_i$ and $\mathbf{V}_t$ are not rank-one matrices, then the following randomized technique introduced in \cite{Wei11} can be utilized to obtain approximated/sub-optimal solution to the original optimization problem in \eqref{proposedDFRC}:
\begin{itemize}
    \item Generate two set of random vectors $\hat{\mathbf{w}}_i \sim \mathcal{CN}(0,\mathbf{F}_i)$ and $\hat{\mathbf{v}}_t\sim \mathcal{CN}(0,\mathbf{V}_t)$;
    \item  Substitute $\mathbf{w}_i=\sqrt{p_i}\hat{\mathbf{w}}_i$ and $\mathbf{v}_t=\sqrt{q_t}\hat{\mathbf{v}}_t$ into \eqref{proposedDFRC} and solve for $p_i$ and $q_t$. This substitution converts the optimization problem in \eqref{proposedDFRC} into a linear programming problem.
    \item Repeat the above two steps several times and select the best solution.
\end{itemize}

Since the randomized technique requires further steps to provide approximated solutions to the original optimization problem \eqref{proposedDFRC}, it not only adds extra computational complexity to the SDR approach but also returns sub-optimal solutions. To tackle the problem, we develop the nature-inspired FA in the following section.
   
\section{Proposed FA Approach}
By letting $\mathbf{z}=\begin{bmatrix}\omega, \textrm{vec}^T\left(\mathbf{W}\mathbf{W}^H+\mathbf{V}\mathbf{V}^H \right) \end{bmatrix}^T$, we can express \eqref{MSE} as:
\begin{eqnarray}
    \mathcal{L}\left( \mathbf{R},\omega\right)=\mathbf{z}^H\mathbf{\Omega}\mathbf{z},
\end{eqnarray}
where $\mathbf{\Omega}$ is defined in \eqref{OmegaMatrix}. The optimization problem in \eqref{proposedDFRC} is equivalently expressed as follows:
\begin{equation} \label{_n}
\begin{aligned}
& \underset{ \{\mathbf{w}_{i} \},\ \{\mathbf{v}_{t} \} }{\textrm{min}} & &
\mathbf{z}^H\mathbf{\Omega}\mathbf{z}\\
& \mbox{s.\ t.}\ & & f_i\left( \{\mathbf{w}_{i}\},\{\mathbf{v}_{t}\}\right) \leq 0, \forall i, \\
&&& d_i\left( \{\mathbf{w}_{i}\},\{\mathbf{v}_{t}\}\right) \leq 0, \ \forall i,\\
&&& g_l(\mathbf{h}_{p,l})\leq 0,\ \forall l,
\end{aligned}
\end{equation}
where 
\begin{eqnarray}
    f_i\left( \{\mathbf{w}_{i}\},\{\mathbf{v}_{t}\}\right)&=&\eta_i\sum_{j=1}^{U}\mathbf{w}_j^H \mathbf{H}_{s,i} \mathbf{w}_j -\left( 1+\eta_i\right)\mathbf{w}_i^H \mathbf{H}_{s,i}\mathbf{w}_i\nonumber\\&{}&+\eta_i\sigma^2_i+\eta_i\sum_{t=1}^{M}\mathbf{v}_t^H \mathbf{H}_{s,i} \mathbf{v}_t,\label{eq_sinr}
\end{eqnarray}

\begin{eqnarray}
    d_i\left( \{\mathbf{w}_{i}\},\{\mathbf{v}_{t}\}\right)&=& \begin{bmatrix}
        \mathbf{W}\mathbf{W}^H+\mathbf{V} \mathbf{V}^H
    \end{bmatrix}_{i,i} -P_m,\nonumber\\
    &=&\mathbf{R}_{i,i}-P_m,\label{eq_Txpower}
\end{eqnarray}
and 
\begin{eqnarray}
    g_l\left( \{\mathbf{w}_{i}\},\{\mathbf{v}_{t}\}\right)=\sum_{j=1}^{U} \mathbf{w}_j^H\mathbf{H}_{p,l}\mathbf{w}_j+\sum_{t=1}^{M}\mathbf{v}_t^H\mathbf{H}_{p,l}\mathbf{v}_t- I_{\text{thres}}.\label{eq_inter}
\end{eqnarray}

In the sequel, we adopt the penalty method in \cite{YangFA2008} to transform the constrained problem \eqref{_n} into the following  equivalent unconstrained problem:
\begin{equation} 
\begin{aligned}
& \underset{ \{\mathbf{w}_{i} \},\ \{\mathbf{v}_{t} \} }{\textrm{min}} & &
\mathbf{z}^H\mathbf{\Omega}\mathbf{z}+ \mathcal{Q}\left(\left\{\mathbf{w}_i\right\},\{\mathbf{v}_{t} \}\right)\label{un_constrained}
\end{aligned}
\end{equation}
where $\mathcal{Q}\left(\left\{\mathbf{w}_i\right\},\{\mathbf{v}_{t} \}\right)$ is the penalty term given as:
    \begin{eqnarray}
    \mathcal{Q}\left(\left\{\mathbf{w}_i\right\},\{\mathbf{v}_{t} \}\right)=\sum_{i=1}^U\lambda_i\text{max}\left\{0, f_i\left(\{\mathbf{w}_i\},\{\mathbf{v}_t\}\right) \right\}^2\label{penalty}\nonumber\\
    +\sum_{i=1}^{M}\rho_i \text{max}\left\{0, d_i\left(\{\mathbf{w}_i\},\{\mathbf{v}_t\}\right)\right\}^2+\sum_{l=1}^L \zeta_l \text{max}\left\{0, g_l\left(\{\mathbf{h}_{p,l}\}\right)\right\}^2,\label{Qformula}
\end{eqnarray}
with $\lambda_i>0$, $\rho_i>0$, and $\zeta_l>0$ being the penalty constants. The values of the penalty constants represent the priority of the respective constraints. A higher value tends to ensure a better accuracy in handling the corresponding constraint.

Note that the first, second, and third constraints of \eqref{_n} are the equivalent forms of the SINR, transmit power, and interference constraints in \eqref{proposedDFRC}. The original SINR, transmit power, and interference constraints are satisfied if the values of \eqref{eq_sinr}, \eqref{eq_Txpower}, and \eqref{eq_inter} are, respectively, less than 0. In this case, it can be easily verified from \eqref{Qformula} that $\mathcal{Q}\left(\left\{\mathbf{w}_i\right\},\{\mathbf{v}_{t} \}\right)=0$. Therefore, optimizing \eqref{un_constrained} will only minimize the original objective function. If any of these constraints are violated, then their corresponding values in \eqref{eq_sinr}, \eqref{eq_Txpower} or \eqref{eq_inter} will be greater than zero, resulting in a positive value of $\mathcal{Q}\left(\left\{\mathbf{w}_i\right\},\{\mathbf{v}_{t} \}\right)$. Consequently, optimizing \eqref{un_constrained} will force both the original objective and the corresponding violating term in $\mathcal{Q}\left(\left\{\mathbf{w}_i\right\},\{\mathbf{v}_{t} \}\right)$ to fall.

The FA was introduced based on the following three rules \cite{YangFA2008,YangFA2009} in a flock of unisex fireflies. First, every firefly attracts the others within the flock. Second, the attractiveness of a firefly to the other firefly in the flock is proportional to its brightness. Third, the landscape of the objective function determines the brightness of a firefly. The brightness and attractiveness increase as the distance between two fireflies decreases. Consider any two fireflies, the darker one will move towards the brighter firefly. A firefly will move randomly when it does not see any brighter firefly. Adopting the generalized FA framework in our earlier work in \cite{Tuan10311527}, we introduce the FA approach to tackle the optimization problem in \eqref{proposedDFRC}.

Let $\left\{\mathbf{W}_m,\mathbf{V}_m, \omega_m \right\}$ be the $m$-th firefly where $\mathbf{W}_m=\begin{bmatrix}\mathbf{w}_1^m, \mathbf{w}_2^m, \cdots, \mathbf{w}_U^m  \end{bmatrix}$ and $\mathbf{V}_m=\begin{bmatrix}\mathbf{v}_1^m, \mathbf{v}_2^m, \cdots, \mathbf{v}_M^m  \end{bmatrix}$. We initialize a flock of $N$ fireflies and express the brightness, a.k.a., the light density, of the $m$-th firefly as: 

\begin{equation}
   B_m\left(\mathbf{W}_m,\mathbf{V}_m, \omega_m\right)=\frac{1}{\mathbf{z}^H\mathbf{\Omega}\mathbf{z}+ \mathcal{Q}\left(\left\{\mathbf{w}_i\right\},\{\mathbf{v}_{t} \}\right)}. \label{lightRIS}
\end{equation}

Consider two fireflies $m$ and $n$ within the flock at pseudo-time $k$. If firefly $m$ is brighter than firefly $n$, i.e., $B_m\left(\mathbf{W}_m,\mathbf{V}_m, \omega_m\right) > B_n\left(\mathbf{W}_n,\mathbf{V}_n, \omega_n\right)$, then firefly $n$ will fly towards firefly $m$ and its variables at iteration $k+1$ are updated as:
\begin{eqnarray}
    \mathbf{W}_n^{(k+1)}=\mathbf{W}_n^{(k)}+\beta_0 e^{- \left(\gamma^{0.5}r_{w,mn}^{(k)}\right)^2}\left(\mathbf{W}_m^{(k)}-\mathbf{W}_n^{(k)} \right)+\alpha^{(k)}\mathbf{W},\label{FAmoveW}\\
    \mathbf{V}_n^{(k+1)}=\mathbf{V}_n^{(k)}+\beta_0 e^{- \left(\gamma^{0.5}r_{v,mn}^{(k)}\right)^2}\left(\mathbf{V}_m^{(k)}-\mathbf{V}_n^{(k)} \right)+\alpha^{(k)}\mathbf{V},\label{FAmoveV}\\
    \omega_n^{(k+1)}=\omega_n^{(k)}+\beta_0 e^{- \left(\gamma^{0.5}r_{\omega,mn}^{(k)}\right)^2}\left(\omega_m^{(k)}-\omega_n^{(k)} \right)+\alpha^{(k)}\omega,\label{FAmoveomega}
\end{eqnarray}
where $r_{w,mn}^{(k)}=|| \mathbf{W}_m^{(k)}-\mathbf{W}_n^{(k)}||$, $r_{v,mn}^{(k)}=|| \mathbf{V}_m^{(t)}-\mathbf{V}_n^{(t)}||$ and $r_{\omega,mn}^{(k)}=|| \omega_m^{(t)}-\omega_n^{(t)}||$ are the Cartesian distances between the two fireflies $m$ and $n$, while, $\beta_0$ is the attractiveness at $r_{w,mn}^{(k)}=0$, $r_{v,mn}^{(k)}=0$ and $r_{\omega,mn}^{(k)}=0$. The attractiveness variation is denoted as $\gamma$. Furthermore, the second terms of \eqref{FAmoveW}, \eqref{FAmoveV}, and \eqref{FAmoveomega} represent the attractions. While, their third terms capture the randomization factor $\alpha^{(t)}$, $\mathbf{W} \in\mathbb{C}^{M \times U} $,   $\mathbf{V} \in\mathbb{C}^{M \times K}$ and $\omega$. The factor $\alpha^{(t)}$, $\omega$, and the elements of $\mathbf{W} $ and   $\mathbf{V} $ are drawn from either a uniform or a Gaussian distribution. 

We proceed by introducing our FA approach as follows. The flock of $N$ fireflies is ranked in descending order of their brightness. The current best solution is declared as the first firefly in the ranked flock. Starting from the first firefly of the sorted flock, each firefly compares its brightness with the current best one and the others. If the firefly finds a brighter one, including itself, than the current best, then the current best will be replaced by that brighter firefly. If the firefly finds a brighter one than itself, it will fly towards that brighter firefly as \eqref{FAmoveW}, \eqref{FAmoveV}, and \eqref{FAmoveomega}. Since the distance between the two fireflies changes after the move, the attractiveness of the brighter one to the current firefly will be updated. Consequently, the brightness of the flock will also be updated, followed by an update ranking of the flock when all fireflies have completed their comparisons with all others. The current best is again declared as the first firefly in the flock. The process is repeated for a predefined maximum number of generations.  The proposed FA approach to solve the optimization problem in \eqref{proposedDFRC} is summarized in Algorithm~\ref{FADFRC}.

The behavior of the firefly population is determined by two main parameters $\gamma$ and $\alpha^{(k)}$. The value of $\gamma$ controls the attractiveness of the fireflies while $\alpha^{(k)}$ determines how the fireflies randomly move. When $\gamma=0$, the attractiveness is a constant with respect to the distance between any firefly. All the fireflies can see each other and can fly toward the brightest one, i.e., the population can reach to the global optimum. When $\gamma \rightarrow \infty$, the attractiveness equals zero, representing the fact that no firefly can be seen by the others. In this case, each firefly randomly moves, hence the global optimum is not always obtained. 

The value of $\gamma$ is normally selected in the range between the above two cases. Therefore, any firefly can be seen by a group of nearby mates where the visibility distance of each group is determined by the characteristic length $\Gamma = $ $\gamma^{-0.5}$. Owning such a nature of population splitting into small groups enables the FA to effectively deal with highly non-linear and multimodal problems \cite{Tuan10311527}. Considering the case when the population size is significantly larger than the number of local optima, the iterations of Algorithm~\ref{FADFRC} will stochastically lead the population to the best solution among the local optima. The fireflies can theoretically attain the global optimum if $k \gg 1$ and $N \rightarrow \infty$ \cite{YangFA2008,YangFA2009,Tuan10311527}. It has been observed in our recent works in \cite{Tuan10311527} and \cite{TuanWCL2024} that the FA can converge within 120 iterations for various transmit beamforming problems.

{\it Remark 1:} In the implementation, the attractiveness between two fireflies is characterized by their Cartesian distance $r$ as
\begin{equation}
\beta\left(r\right)=\beta_0 e^{-\gamma r^m}, m \geq 1,
\end{equation}
which also determines a characteristic length $\Gamma=\gamma^{-1/m}$. The reason is that the proper value of $\gamma r^m$ should be $\mathcal{O}(1)$. This means that $\gamma r^m=1$ gives the characteristic length
$\Gamma=\gamma^{-1/m}$, which is the influence radius of a firefly.
For the Euclidean norm $m=2$, this characteristic length is
$\Gamma=1/\sqrt{\gamma}$. Conversely, for a given scale $\Gamma$
in an optimization, $\gamma$ can be selected as $\gamma=\Gamma^{-m}$,
which is typically in the range from 0.01 to 100  \cite{Yang_FA_Multi_Modal_2009}.

The computational complexity of Algorithm~\ref{FADFRC} is provided in the following lemma.

\begin{lemma}\label{FA_Com_lemma}
The computational complexity of Algorithm~\ref{FADFRC} is dominated by the order of $ \Theta  N^2 G K M^8$.
\end{lemma}

\begin{proof}
See Appendix~\ref{FA_Lemma_Proof}.
\end{proof}

{\it Remark 2:} When the number of antenna elements is large, as in massive MIMO, the entire antenna array can be divided into distinct sub-array groups. In this case, a suitable sub-array of antennas is selected for DFRC transmission instead of utilizing the entire antenna array \cite{10475421,10335685}. Therefore, our proposed approach can still be utilized for the selected sub-array antennas with an affordable computational complexity. While antenna selection methods for DFRC attract increasing research, they are beyond the scope of this work. Interested readers are referred to \cite{10475421,10736998,10335685} and references therein for more details.

\begin{algorithm}
	\caption{Firefly Algorithm for Solving \eqref{proposedDFRC}}
	\label{FADFRC}
	\begin{algorithmic}[1]
	\State  \textbf{Input:} Channel state information $\textbf{h}_{i}, \forall i$; Noise variance $\sigma^2_i$; required SINR $\eta_i$; desired beam pattern $\mathcal{D}\left(\theta_{g,t} \right), \ \forall t$; scaling factor $\omega$; flock size $N$; maximum generation $\Theta$; $\lambda_i$; $\rho_i$; $\zeta_l$, $\beta_0$; $\gamma$;
		\State  Randomly initialize a flock of $N$ fireflies $\left \{ \{\mathbf{W}_1,\mathbf{V}_1,\omega_1\}, \{\mathbf{W}_2,\mathbf{V}_2,\omega_2\},\cdots, \{\mathbf{W}_{N},\mathbf{V}_{N},\omega_{N} \} \right \}$; 
		\State Calculate the brightness of the flock  as \eqref{lightRIS};
		\State  Rank the fireflies in the descending order of $B_m\left(\mathbf{W}_m,\mathbf{V}_m,\omega_m\right)$;
		\State Declare the current best solution: $B^{\star}:=B_1\left(\mathbf{W}_1,\mathbf{V}_1,\omega_1\right)$; $\{\mathbf{W}^{\star},\mathbf{V}^{\star},\omega^{\star}\}:=\{\mathbf{W}_1,\mathbf{V}_1,\omega_1\}$;
		\For{$k =1:\Theta$}
		    \For{$n=1:N$}
		    \For{$m=1:N$}
		    \If{$B_n\left(\mathbf{W}_n,\mathbf{V}_n,\omega_n\right)>B^{\star}$} 
		    \State $B^{\star}:=B_n\left(\mathbf{W}_n,\mathbf{V}_n,\omega_n\right)$; $\{\mathbf{W}^{\star},\mathbf{V}^{\star},\omega^{\star}\}:=\{\mathbf{W}_n,\mathbf{V}_n,\omega_n\}$;
		    \EndIf
		    \If{$B_m\left(\mathbf{W}_m,\mathbf{V}_m,\omega_m\right)>B^{\star}$} 
		    \State $B^{\star}:=B_m\left(\mathbf{W}_m,\mathbf{V}_m,\omega_m\right)$; $\{\mathbf{W}^{\star},\mathbf{V}^{\star},\omega^{\star}\}:=\{\mathbf{W}_m,\mathbf{V}_m,\omega_m\}$;
		    \EndIf
		    \If{$B_m\left(\mathbf{W}_m,\mathbf{V}_m,\omega_m\right)>B_n\left(\mathbf{W}_n,\mathbf{V}_n,\omega_n\right)$}
		\State Move firefly $n$ towards firefly $m$ as \eqref{FAmoveW}, \eqref{FAmoveV}, and \eqref{FAmoveomega};
		    \EndIf
		\State Attractiveness varies with distances via $e^{-\gamma \left(r_{w,mn}^{(k)}\right)^2}$, $e^{-\gamma \left(r_{v,mn}^{(
  k)}\right)^2}$, and $e^{-\gamma \left(r_{\omega,mn}^{(
  k)}\right)^2}$;
		\State Evaluate new solutions and update the brightness of the flock as \eqref{lightRIS};
		    \EndFor 
		    \EndFor 
		    \State Rank the fireflies in a descending order  of $B_m\left(\mathbf{W}_m,\mathbf{V}_m,\omega_m\right)$;
		    \State Update the current best solution: $B^{\star}:=B_1\left(\mathbf{W}_1,\mathbf{V}_1,\omega_1\right)$; $\{\mathbf{W}^{\star},\mathbf{V}^{\star},\omega^{\star}\}:=\{\mathbf{W}_1,\mathbf{V}_1,\omega_1\}$;
		\EndFor
		\State \Return $\mathbf{W}^{\star}, \mathbf{V}^{\star}$.
		\end{algorithmic}
\end{algorithm}
\section{Numerical Results}
\subsection{Simulation Setup}
In this section, we present numerical results to evaluate the performance of the proposed cognitive DFRC with SDR and FA approaches. Here, a BS tracks three targets, i.e., $K=3$, located at $-60^{\circ}$, $0^{\circ}$, and $40^{\circ}$ with respect to the broadside of the BS's array antennas. Hence, the desired beams for the MIMO radar include three main beams $\overline{ \theta }_1=-60^{\circ}$, $\overline{ \theta }_2=0^{\circ}$, and $\overline{ \theta }_3=40^{\circ}$. Each ideal beam has a beam width of $10^{\circ}$. The desired beam patterns $\mathcal{D}$ in \eqref{MSE} for each beam is set as 
\begin{equation} \mathcal{D}(\theta_{g,t}) = \left\lbrace \begin{array}{cl}1, & \overline{ \theta }_t - 5 \leq \theta_{g,t} \leq \overline{ \theta }_t + 5,\\ &\ g = \{1,\cdots,G\},\ t = \{1,2,3\}, \\ 0, & \mathrm{otherwise}. \end{array} \right. 
\end{equation}
The sample angle grids $\{ \theta_{g,t}\}_{g=1}^G$ for each beam pattern are attained by uniformly sampling with a step size of $1^{\circ}$ between $-90^{\circ}$ and $90^{\circ}$, i.e., $G=181$.

The BS serves  two secondary users, i.e., $U=2$, located at $-30^{\circ}$ and $20^{\circ}$ with respect to the broadside of the BS's array antennas. There is one primary user, i.e., $L=1$, located at $-40^{\circ}$ with respect to the broadside of the BS's array antennas.

We adopt the statistical CSI model where the channel
covariance matrices from the BS to the secondary user $i$ , i.e.,
$\mathbf{H}_{s,i}=\mathbf{H}\left( \zeta_{s,i}\right)$, and to primary $l$, i.e., $\mathbf{H}_{p,l}=\mathbf{H}\left( \zeta_{p,l}\right)$, are the function of the angle of departure, i.e., $\zeta_{s,i}$ or $\zeta_{p,l}$. The $(m,n)$-th element of the channel
covariance matrix $\mathbf{H}\left(
\zeta\right)$ is, \cite{Mats,Tuanglobecom11}:
\begin{equation} \label{covaformula}
\mathbf{H}\left(
\zeta\right)_{(m,n)}=e^{\frac{j2\pi \Delta}{\psi}\left[\left(n-m\right)\text{sin}\zeta\right]} e^{-2\left[\frac{\pi \Delta \delta_a}{\psi}\left\{\left(n-m\right)\text{cos}\zeta\right\}\right]^2},
\end{equation}
 where $\psi$ is the carrier wavelength, $\delta_a=2^\circ$ is the standard deviation of the angular spread, and the antenna
spacing at the BS is set as $\Delta=\psi/2$. It is noted that, without loss of generality, the wireless channel in our experiments is normalized to exclude the deterministic effect, i.e., pathloss and real noise power spectral density, \cite{Xiang2020}. Similar to the other works, e.g., \cite{Yongwei,Xiang2020}, the noise variance is set to $\sigma^2=0.1$ while the interference tolerance threshold is $I_{\text{thres}}=0.01$ which is $10$ dB lower than the noise level. The BS's transmit power budget $P_m$ is set to either 1, 1.25, or 1.5, which is $10$ dB, $10.97$ dB, and $ 11.77$ dB higher than the noise level, respectively.

CVX package\footnote{CVX is a modeling system for constructing and solving disciplined convex programs (DCPs) where a set of conventions or rules, a.k.a., DCP ruleset, is imposed. Problems that adhere to the ruleset can be rapidly and automatically verified as convex and converted to solvable form \cite{cvx2015}. CVX supports several standard problem types, including linear and quadratic programs (LPs/QPs), second-order cone programs (SOCPs), and semidefinite programs (SDPs). CVX allows users to declare standard problems in their closest natural mathematical forms and exploit standard solvers, i.e., interior point methods, to find optimal solutions. }  is adopted to obtain the solution for the SDR approach. The setup parameters for the FA approach are as follows. Since the optimization variables in our implementation are well-scaled, i.e., neither too small nor too large, the value of the attractiveness variation $\gamma$ is set to 1. The penalty constants are set equal, but they dynamically update as $\lambda_i=\rho_i= \zeta_l=k^2, \ \forall i, l$, where $k$ is the iteration index in Algorithm~\ref{FADFRC}. As the iterations of the FA algorithm increase, a higher priority is put toward the constraints to ensure they are strictly satisfied.\footnote{Although unequal penalty constants are possible, a uniform penalty constant is used for all constraints in this work as they are treated with equal importance.} The attractiveness at zero distance is set as $\beta_0=1$ to maintain the scale of the firefly's moves defined in \eqref{FAmoveW}, \eqref{FAmoveV}, and \eqref{FAmoveomega}. Finally, the initial randomization factor is $\alpha^{(0)}=0.9$  and its value at the $k$-th generation is $\alpha^{(k)}=\alpha^{(0)}0.93^k$. The selection of the initial randomization factor and its dynamic update is to ensure that when Algorithm~\ref{FADFRC} starts, the FA maintains a highly random walk property, a.k.a, the exploration ability, to avoid the fireflies being trapped at local optimals. As the iterations progress, the randomness factor decreases, hence, the fireflies can move quicker to the brightest ones. The population size is either $N=40$ or $N=60$ and the maximum number of generations is $\Theta=140$.

\subsection{Radiation Beam Pattern Comparisons}
\begin{figure}[t]
\centering
    \includegraphics[width=.45\textwidth]{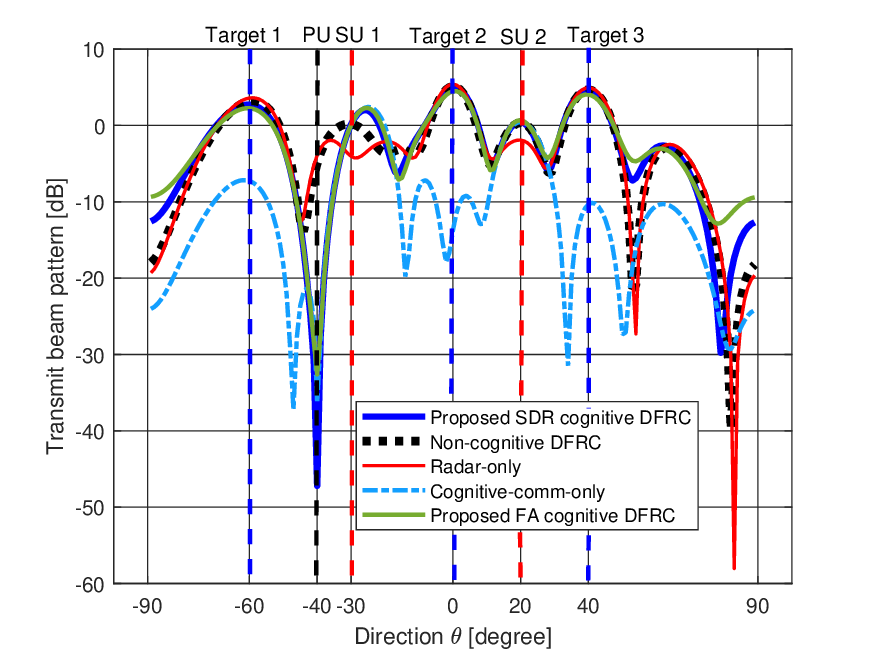}
\caption{\label{10anBeamsComp} Radiation beam
patterns of the BS with $M=10$ antennas for different approaches. The interference tolerance level at the PU is $I_{\text{thres}}=0.01$. The required SINR at each SU is $\eta_i=10$ dB. The FA population size is $N=40$. The number of FA generations is $\Theta=140$. $P_m=1$.}
\end{figure}

\begin{table*}[ht]
  \centering
  	\begin{tabular}{ |l|c|c|c| } 
		\hline
		 & \textbf{Radar Function} & \textbf{Communication Function} & \textbf{Cognitive Function}  \\
		
		\hline
	    Proposed SDR cognitive approach & Yes& Yes & Yes\\
	    \hline
	    Proposed FA cognitive approach & Yes& Yes & Yes\\
	    \hline
	Non-cognitive DFRC approach \cite{Xiang2020} & Yes& Yes & No\\
	    \hline
	Radar-only approach \cite{Stoica} & Yes& No& No\\
	    \hline
        Cognitive-comm-only approach \cite{Yongwei} & No& Yes & Yes\\
	    \hline
	\end{tabular}
  \caption{Function Comparisons}
  \label{tab:1}
\end{table*}
Fig.~\ref{10anBeamsComp} compares the transmit beam patterns of the proposed cognitive DFRC approaches, i.e., SDR and FA approaches, against those of the non-cognitive DFRC approach, i.e., problem (31) of  \cite{Xiang2020}, the radar-only approach, i.e., problem (12) of \cite{Stoica}, and the cognitive-comm-only approach, i.e., problem (12) of \cite{Yongwei}.\footnote{For a fair comparison,  problems (31) of \cite{Xiang2020} and (12) of \cite{Stoica} have the same objective function as that of the proposed problem \eqref{proposedDFRC}.} The radar-only approach steers its probing beams towards 3 targets without caring for the communication users, i.e., the primary and secondary users. The cognitive-comm-only approach serves 2 secondary users while imposing an acceptable interference level on the primary user. It neither cares nor knows of the existence of the targets. The non-cognitive DFRC approach tracks 3 targets while communicating with 2 secondary users. However, it does not take care of the primary user. On the other hand, the proposed SDR and FA cognitive DFRC approaches simultaneously communicate with 2 secondary users and track 3 targets while putting an acceptable interference level on the primary user. The comparisons of functions between the proposed and baseline approaches are shown in Table~\ref{tab:1}.

Due to their single-purpose designs, the radar-only and cognitive-comm-only approaches require less computational complexities than the proposed SDR and FA cognitive DFRC counterparts. However, the main drawback of the single-purpose approaches is inefficient resource utilization, e.g., spectrum and hardware. Non-cognitive DFRC approach is a jointly designed radar and communication system, yet it is not equipped with the cognitive ability. Thus, the non-cognitive DFRC approach utilizes spectrum less effectively in return for requiring less computational complexity than the proposed SDR and FA cognitive DFRC approaches.

 It is clear from Fig.~\ref{10anBeamsComp} that the FA approach yields almost identical beam patterns as the CVX solver does for the SDR approach. Next, we verify the transmit beam patterns of five approaches against three functions shown in Table~\ref{tab:1}. {\it i) Radar function:} It can be seen from the figure that four approaches, i.e., proposed SDR, FA cognitive DFRC, non-cognitive DFRC, and the radar-only approaches, steer the same beams towards the three targets, i.e., at $-60^{\circ}$, $0^{\circ}$, and $40^{\circ}$ while the cognitive-comm-only approach does not. {\it ii) Communication function: } Note that the noise level is set at $0.1$ or $-10$ dB. So with the required SINR level of 10 dB, we are expecting the beam levels at the locations of secondary users 1 and 2, i.e., at $-30^{\circ}$ and $20^{\circ}$, are above 0 dB. It is clear that the proposed SDR, FA cognitive DFRC, cognitive-communication-only, and non-cognitive DFRC approaches satisfy the requirement. On the other hand, the radar-only approach has much lower beam levels at these directions as expected since it does not have the communication function. {\it iii) Cognitive function: } The interference threshold accepted by the primary user is $I_{\text{thres}}=0.01$ or -20 dB. It can be observed from the figure that the proposed SDR, FA cognitive DFRC and cognitive-communication-only approaches effectively form a deep null, i.e., much lower than -20 dB, to protect a primary user at $-40^{\circ}$ whereas the non-cognitive DFRC and radar-only approach fail to maintain an acceptable interference level, i.e., $-20$ dB, at the primary user.

\begin{figure}[t]
\centering
    \includegraphics[width=.45\textwidth]{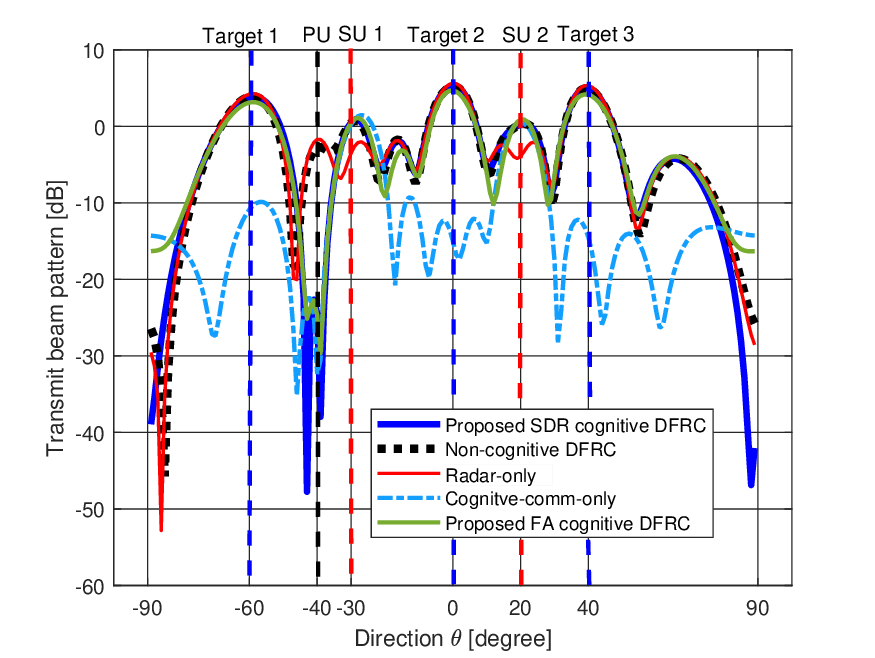}
\caption{\label{12anBeamsComp} Radiation beam
patterns of the BS with $M=12$ antennas for different approaches. The interference tolerance level at the PU is $I_{\text{thres}}=0.01$.  The required SINR at each SU is $\eta_i=10$ dB. The FA population size is $N=40$. The number of FA generations is $\Theta=140$. $P_m=1$.}
\end{figure}

\begin{figure}[t]
\centering
    \includegraphics[width=.45\textwidth]{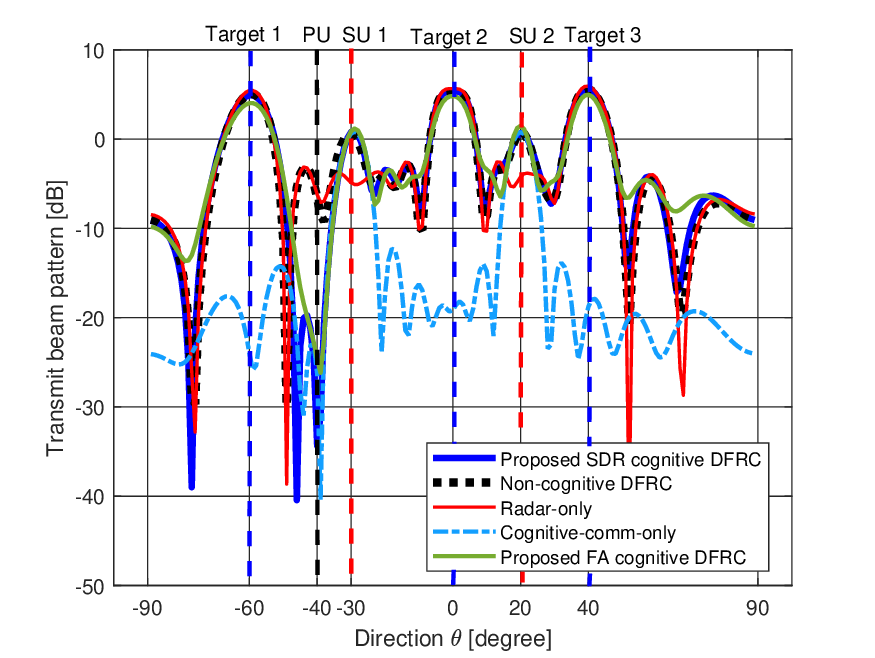}
\caption{\label{16anBeamsComp} Radiation beam
patterns of the BS with $M=16$ antennas for different approaches. The interference tolerance level at the PU is $I_{\text{thres}}=0.01$. The required SINR at each SU is $\eta_i=10$ dB. The FA population size is $N=40$. The number of FA generations is $\Theta=140$. $P_m=1$.}
\end{figure}

Similar trends in Fig.~\ref{10anBeamsComp} can also been observed in Figs.~\ref{12anBeamsComp} and \ref{16anBeamsComp} where those the transmit beam patterns are shown for $M=12$ and $M=16$, respectively. However, as the number of antennas increases, the resolution of the MIMO antenna array improves which results in sharper and more effective beams. This, in turn, improve the system performance which will be seen and discussed later.

\subsection{Convergence of the Proposed FA Cognitive DFRC Approach}
\begin{figure}[t]
\centering
    \includegraphics[width=.45\textwidth]{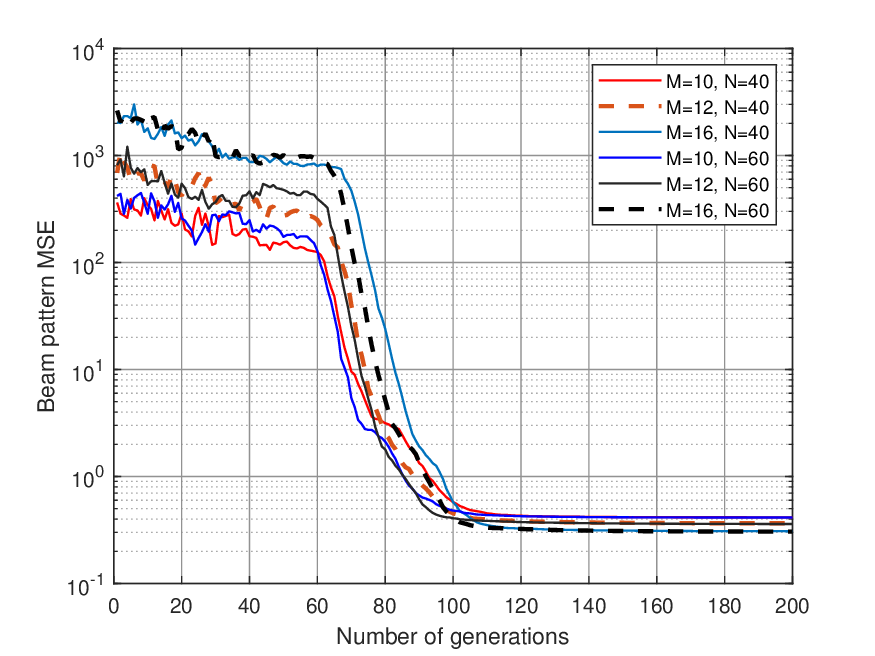}
\caption{\label{FA_MSE_vs_G} Beam
pattern MSE of the proposed FA approach versus the number of generations/iterations for different number of antenna elements, $M$ and population size, $N$. The interference tolerance level at the PU is $I_{\text{thres}}=0.01$. The required SINR at each SU is $\eta_i=10$ dB. $P_m=1$.}
\end{figure}

The convergence of the proposed FA approach is shown in Fig.~\ref{FA_MSE_vs_G}. The figure illustrates the beam pattern MSE versus the number of generations used in Algorithm \ref{FADFRC} with different numbers of antenna elements $M$ and firefly population size $N$. It is clear from the figure that the proposed FA approach converges after $110$ generations for all observed settings of $M$ and $N$. 
\begin{figure}[t]
\centering
    \includegraphics[width=.45\textwidth]{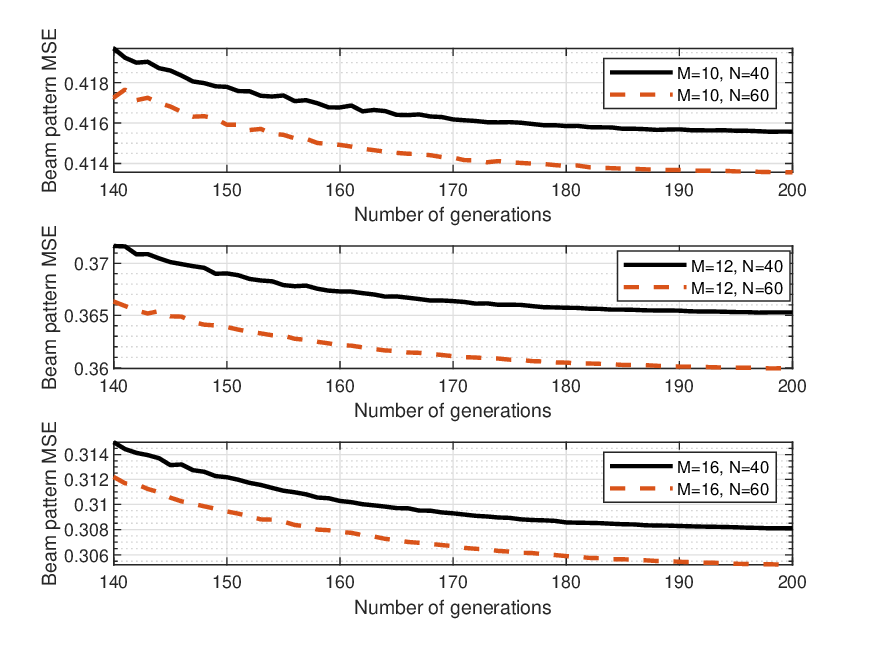}
\caption{\label{FA_MSE_vs_G_zoomin} Beam
pattern MSE of the proposed FA approach versus the number of generations/iterations for different number of antenna elements, $M$ and population size, $N$. The interference tolerance level at the PU is $I_{\text{thres}}=0.01$. The required SINR at each SU is $\eta_i=10$ dB. $P_m=1$.}
\end{figure}

Fig.~\ref{FA_MSE_vs_G_zoomin} zooms in the convergence behavior of the proposed FA shown in Fig.~\ref{FA_MSE_vs_G} from generation $140$ to $200$. The results reveal that when the number of generations increases from $140$ to $170$, the MSE can be further reduced by $0.003$ with $M=10$ and by $0.005$ with $M=12$ and $M=16$. On the other hand, increasing the population size from $N=40$ to $N=60$ can improve the MSE performance of the proposed FA approach by $0.002$ with $M=10$ and by $0.003$ with $M=12$ and $M=16$ at the number of generations of $160$.
\subsection{Beam Pattern MSE Comparisons}
\begin{figure}[t]
\centering
    \includegraphics[width=.45\textwidth]{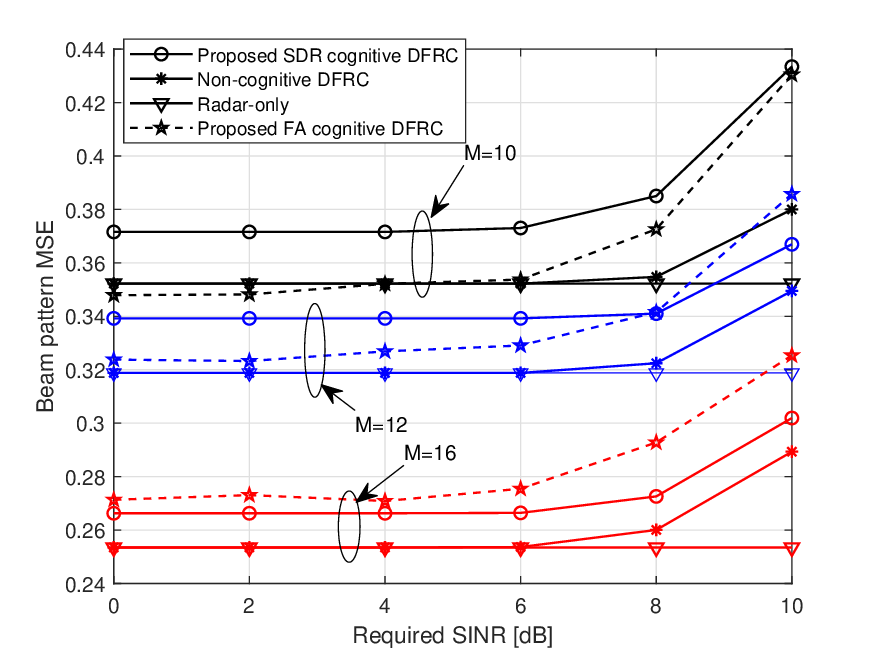}
\caption{\label{MSE_vs_SINR} Beam
pattern MSE versus the target SINR level at secondary users for different approaches with different numbers of antennas, $M$. The interference tolerance level at the PU is $I_{\text{thres}}=0.01$. The FA population size is $N=40$. The number of FA generations is $\Theta=140$. $P_m=1$.}
\end{figure}

Fig.~\ref{MSE_vs_SINR} plots the beam pattern MSE of the proposed SDR cognitive DFRC, proposed FA cognitive DFRC, non-cognitive DFRC, and radar-only approaches versus the required SINR level with different numbers of antennas at the BS.\footnote{The cognitive-comm-only approach is excluded because it does not have the radar function.}  It can be seen that the radar-only approach attains the lowest MSE in the observed SINR range. The non-cognitive DFRC approach follows the same MSE performance at that of the radar-only approach in the SINR range from $0$ dB to $8$ dB.  At SINR of $10$ dB, the non-cognitive approach has higher MSE than the radar-only counterpart does. This is due to the fact that the radar function has been traded off for the communication function in order to provide a relatively high SINR level whereas in lower SINR range, the communication function can be met without any loss of the radar function.

Fig.~\ref{MSE_vs_SINR} indicates that maintaining the required null at the PU while transmitting to SUs results in the highest MSE level. For instance the MSE offered by the proposed SDR cognitive DFRC is $0.2$ higher than that offered by the non-cognitive DFRC in the SINR range from $0$ dB to $6$ dB with $M=10$ and $M=12$. The gap is reduced to $0.1$ with $M=16$. This performance gap is because of the fact that having more constraints on the optimization problem narrows down the feasibility set which in turn affects on the optimality. The results shown on Fig.~\ref{MSE_vs_SINR} also reveal a fact that the MSE can be reduced when more transmit antennas are available. For example, the MSE attained by the proposed SDR coginitve DFRC is reduced from $0.373$ to $0.266$ at the SINR level of $6$ dB when the number of antenna increases from 10 to 16. The improvement of MSE is due to the benefit of having more degrees of freedom from the increment of number of antennas, i.e., improving the antenna resolution.

It can be seen from Fig.~\ref{MSE_vs_SINR} that with the settings of $M=10$ and $M=12$, the proposed FA approach attains lower MSE than the proposed SDR approach does in the SINR range from $0$ dB to $6$ dB. For instance, at SINR level of $2$ dB, the MSE gaps between the SDR and the FA are, respectively, $0.03$ and $0.02$ with $M=10$ and $M=12$. When the required SINR is beyond, $6$ dB, the performance of the FA is closed with that of the SDR with $M=10$ while its MSE is $0.02$ higher than that of the SDR at $10$ dB with $M=12$. 

With $M=16$, the proposed SDR obtains $0.005$  lower MSE than that of the proposed FA in the SINR range from $0$ dB to $6$ dB. From $8$ dB to $10$ dB the MSE of the FA is $0.02$ higher than that of the SDR. However, as shown in Fig.~\ref{FA_MSE_vs_G_zoomin}, the performance of the FA can closely follow that of the SDR counterpart with $60$ fireflies at $190$ generations. This is due to a fact that 
the size of the proposed optimization problem increases with a higher number of antenna elements. Hence, more fireflies are required to enhance the diversification for the FA's exploration.

\begin{figure}[t]
\centering
    \includegraphics[width=.45\textwidth]{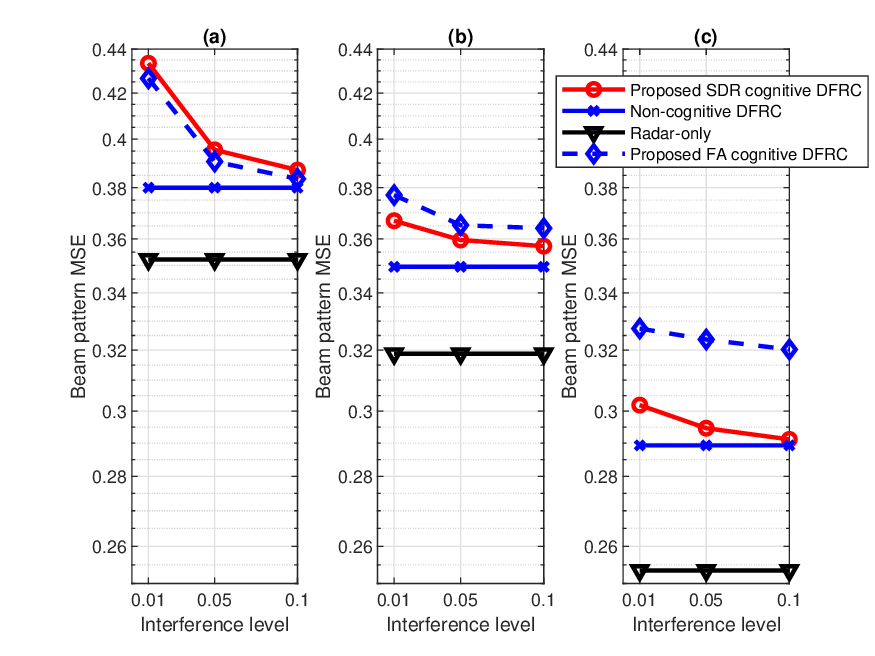}
\caption{\label{MSE_vs_inter} Beam
pattern MSE versus the interference tolerance level for different approaches with the number of antennas:  a) $M=10$, b) $M=12$, c) $M=16$. The required SINR at each SU is $\eta_i=10$ dB. $P_m=1$.}
\end{figure}

In Fig.~\ref{MSE_vs_inter}, the beam pattern MSE is plotted versus the interference tolerance level at PU for the proposed SDR and FA cognitive DFRC approaches, the non-cognitive DFRC approach and the radar-only approach. The required SINR level at each SU is $10$ dB. As expected, the non-cognitive and radar-only approaches have a constant level of MSE as they do not take into account the existence of the PU. It can be observed from the figure that adding more functions, i.e., communication function and cognitive function, onto the system results in a higher beam pattern MSE level. This is the cost of using frequency resource from the primary system. The beam pattern MSE of the proposed SDR cognitive DFRC approach converges to that of the non-cognitive DFRC method as the interference tolerance level increases. Intuitively, when a strict interference tolerance level is required, i.e., a low level of interference, the proposed SDR cognitive DFRC is desirable. On the other hand, when the tolerance level is not strict, i.e., the primary system accepts high level of interference, the non-cognitive DFRC can operate. Fig.~\ref{MSE_vs_inter} illustrates that the proposed FA cognitive DFRC attains a lower MSE than the proposed SDR cognitive DFRC by around $0.004$ to $0.007$ at $10$ antennas. However, at $12$ antennas, the MSE of the FA is higher than the SDR by around $0.006$ to $0.009$. With $16$ antennas, this difference increases further to around $0.025$ to $0.03$. A similar effect of the number of antenna on the MSE shown in Fig.~\ref{MSE_vs_SINR} can also be observed in Fig.~\ref{MSE_vs_inter}. 

\begin{figure}[t]
\centering
    \includegraphics[width=.45\textwidth]{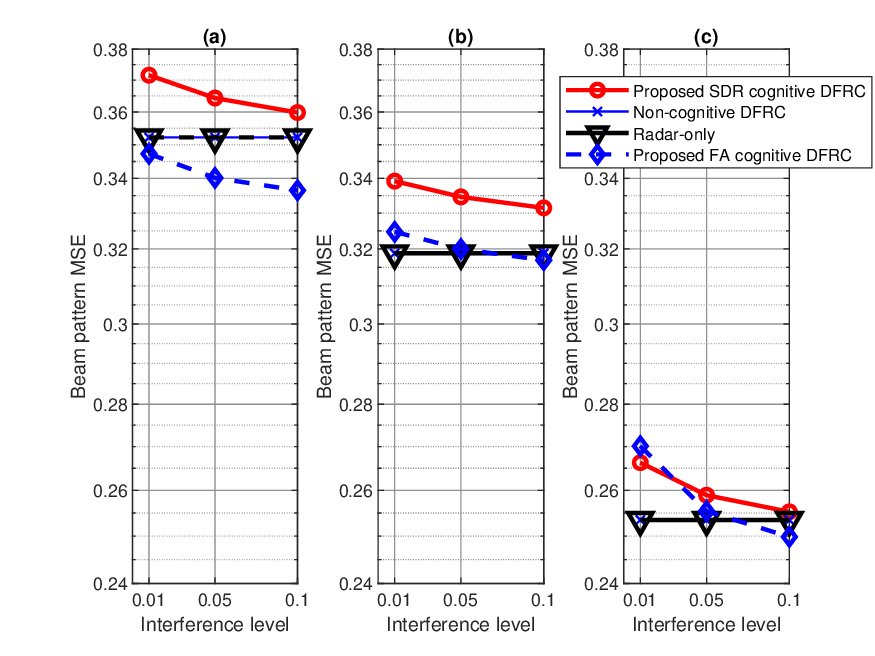}
\caption{\label{MSE_vs_inter4db} Beam
pattern MSE versus the interference tolerance level for different approaches with the number of antennas:  a) $M=10$, b) $M=12$, c) $M=16$. The required SINR at each SU is $\eta_i=4$ dB. $P_m=1$.}
\end{figure}

Fig.~\ref{MSE_vs_inter4db} shows the beam pattern MSE versus the interference tolerance level with the required SINR level of $4$ dB. The result observed in Fig.~\ref{MSE_vs_SINR}, i.e., at relatively low SINR level the DFRC can manage both communication and radar functions without any loss in the performance of the radar function, is confirmed here again with the interference tolerance level range from $0.01$ to $0.1$ where the MSE performance of the non-cognitive DFRC is the same as that of the radar-only. The MSE performance trend of the proposed SDR cognitive DFRC approach at a lower SINR level, i.e., $4$ dB, is similar to that at a higher SINR level, i.e., $10$ dB, shown in Fig.~\ref{MSE_vs_inter}.

Fig.~\ref{MSE_vs_inter4db}~(a) reveals the fact that the proposed FA cognitive DFRC outperforms other approaches in providing lowest beam pattern MSE with 10 antenna elements. As seen in Fig.~\ref{MSE_vs_inter4db}~(b), when the number of antenna elements increases to 12, the performance of the proposed FA is almost the same as those of the non-cognitive DFRC and radar only approaches. With 16 antenna elements, i.e., Fig.~\ref{MSE_vs_inter4db}~(c), the MSE of the proposed FA is slightly higher than that of the proposed SDR at the interference tolerance level of $0.01$. However, the MSE of the proposed FA closely reaches those of the non-cognitive and radar only approaches when the interference tolerance level increases to $0.05$. It then attains the lowest MSE level amongst all approaches at the interference tolerance of $0.1$. 

The behavior of the proposed FA cognitive DFRC shown in Figs.~\ref{MSE_vs_inter} and \ref{MSE_vs_inter4db} indicates a fact that a proper setup of FA parameters, i.e., the population size and maximum number of generations, will well capture the optimization problem in terms of size, i.e., the number of antenna elements, and feasibility region, i.e., the interference tolerance constraint and the SINR constraint. As a result, the FA can operate at its best abilities in exploration and exploitation to offer the best solution.

\begin{figure}[t]
\centering
    \includegraphics[width=.45\textwidth]{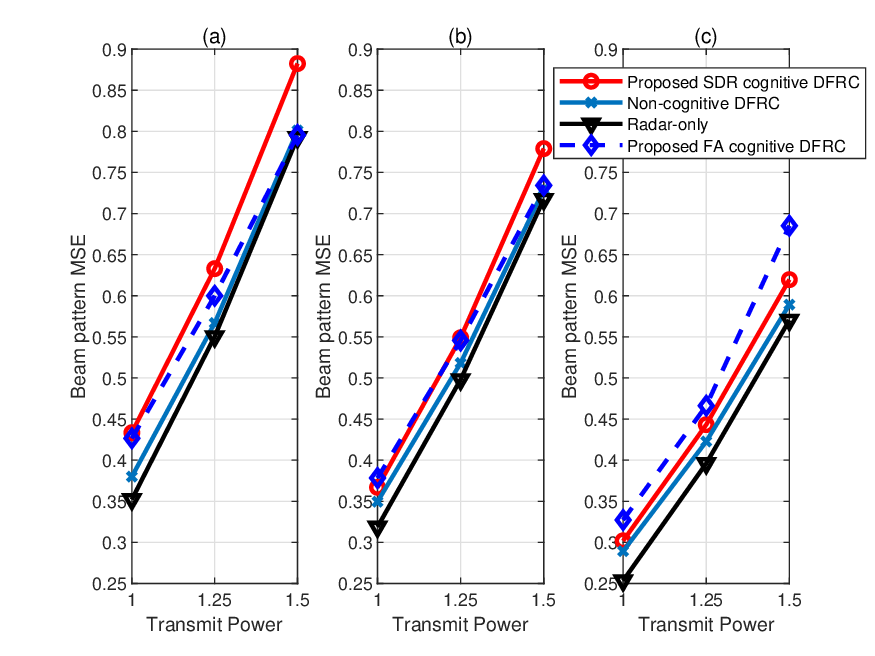}
\caption{\label{MSE_vs_Tx} Beam
pattern MSE versus transmit power for different approaches with the number of antennas:  a) $M=10$, b) $M=12$, c) $M=16$. The required SINR at each SU is $\eta_i=10$ dB.}
\end{figure}

\begin{figure}[t]
\centering
    \includegraphics[width=.45\textwidth]{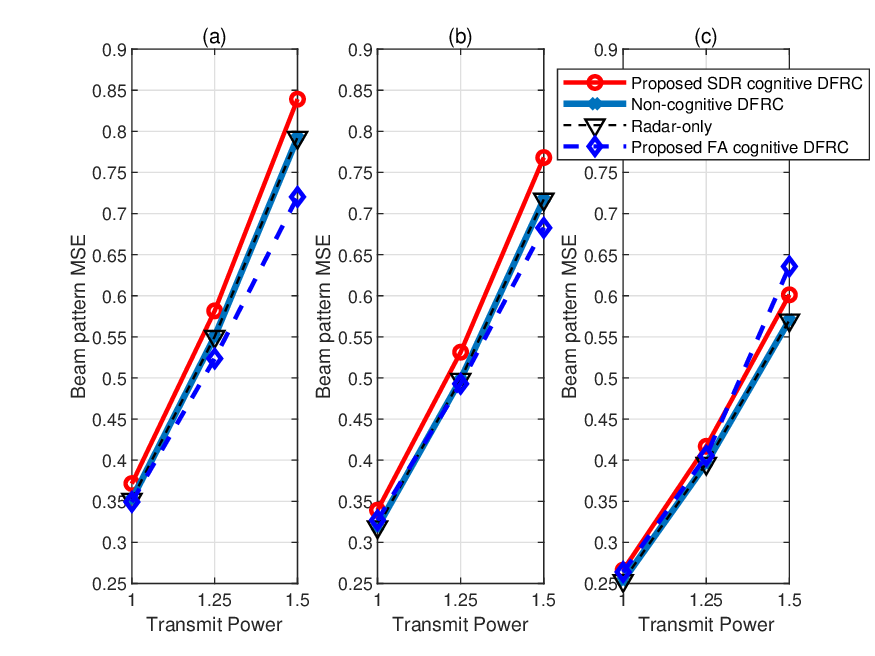}
\caption{\label{MSE_vs_Tx4dB} Beam
pattern MSE versus transmit power for different approaches with the number of antennas:  a) $M=10$, b) $M=12$, c) $M=16$. The required SINR at each SU is $\eta_i=4$ dB.}
\end{figure}

Figs.~\ref{MSE_vs_Tx} and \ref{MSE_vs_Tx4dB} depict the beam pattern MSE versus the transmit power $P_m$ with the required SINR of 10 dB and 4 dB, respectively. Interestingly, as the transmit power budget increases, the performances of all observed approaches decrease, i.e., the beam pattern MSE increases. This is due to the fact that the increased transmit power results in higher peaks towards the radar targets. Unfortunately, their sidelobes also increase, degrading the MSE performance. It can be also seen from Figs.~\ref{MSE_vs_Tx} and \ref{MSE_vs_Tx4dB} that using more antenna elements can provide higher array resolutions hence can effectively manage the sideslobes to reduce the beam pattern MSE. Last but not least, the behaviors of the proposed SDR and FA cognitive DFRC approaches in comparison with their baseline counterparts shown in Figs.~\ref{MSE_vs_Tx} and \ref{MSE_vs_Tx4dB}  are similar to those observed and discussed earlier in Figs.~\ref{MSE_vs_inter} and \ref{MSE_vs_inter4db}.

\section{Conclusion}
This paper proposed an optimization problem for a cognitive DFRC system. Subsequently, the SDR and nature-inspired FA approaches have been developed to attain solutions to the optimization problem. Simulation results confirmed the effectiveness of the joint beamforming design to simultaneously provide communication and radar functions under a spectrum sharing scenario. Specifically, the proposed SDR and FA approaches prevail their baseline counterparts in providing deep nulls towards primary users while effectively communicating with their secondary users and tracking different radar targets. Such ability of exploiting licensed spectrum resource comes at the cost of an increasing beam pattern MSE.  Considering two proposed approaches, the FA outperforms its SDR counterpart in attaining lower beam pattern MSE when the number of antenna elements are 10 and 12 while the required SINR level is at less than 8 dB. Conversely, when the number of antenna elements increases to 16, the SDR attains a lower beam pattern MSE as compared to the FA approach. To compensate this performance gap for large number of antenna elements, the firefly population size of the FA approach needs to be increased in order to closely match the SDR's performance, resulting in a higher computational cost. Further investigations on tackling imperfect CSI will complement this work.

\appendices

\section{Proof of Lemma~\ref{SDR_Com_lemma}}\label{SDR_proof}
Using the barrier method the inequality constraints of problem \eqref{proposedDFRC} can be incorporated into the objective as follows:
\begin{equation}
\begin{aligned}\label{prob_ISAC_new}
& \underset{  \mathbf{F}_i, \ \mathbf{V}_t}{\textrm{min}}  & & \mathbf{r}^H\mathbf{\Omega}\mathbf{r} - \frac{1}{\tau} B\left(\mathbf{F}_i, \mathbf{V}_t \right),
\end{aligned}
\end{equation}
where $\tau$ is the barrier parameter, i.e., a large value of $\tau$ pushes the solution of the above problem to the edge of the feasibility region, and $ B\left(\mathbf{F}_i, \mathbf{V}_t \right)$ is the barrier function given as
\begin{eqnarray}
    B\left(\mathbf{F}_i, \mathbf{V}_t \right)=\sum_{i=1}^U\log \left(\left(1+\frac{1}{\eta_i} \right) \textrm{Tr}\left(\mathbf{H}_{s,i}^H\mathbf{F}_{i}\right)\right.\ \nonumber \\ \left.\ -\sum_{j=1}^U\textrm{Tr}
\left(\mathbf{H}_{s,i}^H\mathbf{F}_{j}\right) -\sum_{t=1}^M\textrm{Tr}
\left(\mathbf{H}_{s,i}^H\mathbf{V}_{t}\right)-\sigma^2_i \right)  \nonumber\\
-\log \sum_{i=1}^{U}\left( \mathbf{R}_{i,i}-\frac{P_m}{M}\right)\nonumber\\
+\sum_{l=1}^K\log\left( I_{\text{thres}}- \sum_{j=1}^U\textrm{Tr}\left( \mathbf{H}_{p,l} \mathbf{F}_{j}  \right)-\sum_{t=1}^M\textrm{Tr}\left( \mathbf{H}_{p,l} \mathbf{V}_{t}  \right)\right)\nonumber\\
-\sum_{i=1}^U \log\left( \det \mathbf{F}_i^{-1} \right)-\sum_{t=1}^M \log\left( \det \mathbf{V}_t^{-1} \right).
\end{eqnarray}

Since \eqref{prob_ISAC} has $(2U+M+L)$ linear matrix inequalities, according to \cite[Chapter 11]{Boyd_convex}, the best bounds on the number of Newton steps for solving \eqref{prob_ISAC_new} based on the barrier approach is on the order of $ \sqrt{(U+M+L)}$. In each Newton step of the barrier method, the complexity is dominated by the  complexity is dominated by the calculation of the Hessian of \eqref{prob_ISAC_new} which includes the Hessian evaluations of $\mathbf{r}^H\mathbf{\Omega}\mathbf{r}$ and $ B\left(\mathbf{F}_i, \mathbf{V}_t \right)$.

It can be observed that the Hessian of original objective $\mathbf{r}^H\mathbf{\Omega}\mathbf{r}$ is $2 \mathbf{\Omega}$. Therefore, the computation complexity to calculate the Hessian is on the same order as that of evaluating $\mathbf{\Omega}$ which is on the order of $GKM^8$. On the other hand, the computational complexity of the Hessian of the barrier function $ B\left(\mathbf{F}_i, \mathbf{V}_t \right)$ is dominated by the order of $M^8$. Therefore, the computational complexity in each Newton step is dominated by the order of $GKM^8$. Consequently, the complexity of the barrier method is on the order of $ \sqrt{(U+M+L)} GKM^8$ which is dominated by $GKM^{8.5}$ when considering the fact that $U+L<M$.

\section{Proof of Lemma~\ref{FA_Com_lemma}}\label{FA_Lemma_Proof}
The proof is based on the following observations. The dominant terms of the computational complexity of Algorithm~\ref{FADFRC} are to generate $N$ fireflies, to evaluate the brightness of the firefly flock, to rank $N$ fireflies, and to move the fireflies. The complexity of generating $N$ fireflies in is on the order of $N M(M+U)$. The complexities of evaluating  $f_i\left( \{\mathbf{w}_{i}\},\{\mathbf{v}_{t}\}\right)$, 
$d_i\left( \{\mathbf{w}_{i}\},\{\mathbf{v}_{t}\}\right)$, and $g_i\left( \{\mathbf{w}_{i}\},\{\mathbf{v}_{t}\}\right)$ are on the order of $M^2(M+U)$. Hence, the complexity of calculating $\mathcal{Q}$ in \eqref{Qformula} is on the order of $M^2(M+U)(M+U+L)$. The complexity of evaluating the objective function is on the order of $GKM^8$. The complexity of evaluating the light density for $N$ fireflies is on the order of  $N\left(GKM^8+M^2(M+U)(M+U+L)\right)$. The complexity of ranking $N$ firefly is $N\log{N}$. Finally, the complexity of moving a firefly in step 16 is on the order of $M(M+U)$. Assuming the worst case when every firefly moves in every inner loop of the algorithm, one can derive the computational complexity of Algorithm~\ref{FADFRC}, after some manipulations, as on the order of:
\begin{eqnarray}
    &{}&\Theta N \log{N}+\Theta N^2 M(M+U)\nonumber\\
    &{}&+\Theta N^2\left(GKM^8+M^2(M+U)(M+U+L)\right).\label{Comp_FARIS}
\end{eqnarray}
Since $U+K+L \leq M$, the complexity of Algorithm~\ref{FADFRC} will be dominated by $\Theta  N^2 G K M^8$.
\bibliographystyle{IEEEtranTCOM}

\begin{thebibliography}{10}

\bibitem{Keivan2010}
M.~G. Khoshkholgh, K.~Navaie, and H.~Yanikomeroglu, ``Access strategies for spectrum sharing in fading environment: Overlay, underlay and mixed,'' \emph{IEEE Trans. Mobile Comput.}, vol.~9, no.~12, pp. 1780--1793, Dec. 2010.

\bibitem{TuanTcom2014}
T.~A. Le and K.~Navaie, ``Downlink beamforming in underlay cognitive cellular networks,'' \emph{IEEE Trans. Commun.}, vol.~62, no.~7, pp. 2212--2223, Jul. 2014.

\bibitem{Zhang10005142}
T.~Zhang, G.~Li, S.~Wang, G.~Zhu, G.~Chen, and R.~Wang, ``{ISAC}-accelerated edge intelligence: Framework, optimization, and analysis,'' \emph{IEEE Trans. Green Commun. and Netw.}, vol.~7, no.~1, pp. 455--468, Mar. 2023.

\bibitem{Liu8386661}
F.~Liu, L.~Zhou, C.~Masouros, A.~Li, W.~Luo, and A.~Petropulu, ``Toward dual-functional radar-communication systems: Optimal waveform design,'' \emph{IEEE Trans. Signal Process.}, vol.~66, no.~16, pp. 4264--4279, Aug. 2018.

\bibitem{Xiang2020}
X.~Liu, T.~Huang, N.~Shlezinger, Y.~Liu, J.~Zhou, and Y.~C. Eldar, ``Joint transmit beamforming for multiuser {MIMO} communications and {MIMO} radar,'' \emph{IEEE Trans. Signal Process.}, vol.~68, pp. 3929--3944, Jul. 2020.

\bibitem{Liu9737357}
F.~Liu, Y.~Cui, C.~Masouros, J.~Xu, T.~X. Han, Y.~C. Eldar, and S.~Buzzi, ``Integrated sensing and communications: Toward dual-functional wireless networks for {6G and Beyond},'' \emph{IEEE J. Sel. Areas in Commun.}, vol.~40, no.~6, pp. 1728--1767, Jun. 2022.

\bibitem{1stDFRC}
R.~M. Mealey, ``A method for calculating error probabilities in a radar communication system,'' \emph{IEEE Trans. Space Electron. and Telemetry}, vol.~9, no.~2, pp. 37--42, Jun. 1963.

\bibitem{HassanienIET}
A.~Hassanien, M.~G. Amin, Y.~D. Zhang, and F.~Ahmad, ``Phase-modulation based dual-function radar-communications,'' \emph{IEEE Radar, Sonar \& Navigation}, vol.~10, no.~8, p. 1411–1421, 2016.

\bibitem{Hassanien7347464}
------, ``Dual-function radar-communications: Information embedding using sidelobe control and waveform diversity,'' \emph{IEEE Trans. Signal Process.}, vol.~64, no.~8, pp. 2168--2181, Apr. 2016.

\bibitem{Wang8438940}
X.~Wang, A.~Hassanien, and M.~G. Amin, ``Dual-function {MIMO} radar communications system design via sparse array optimization,'' \emph{IEEE Trans. on Aerosp. Electron. Syst.}, vol.~55, no.~3, pp. 1213--1226, Jun. 2019.

\bibitem{Sturm5776640}
C.~Sturm and W.~Wiesbeck, ``Waveform design and signal processing aspects for fusion of wireless communications and radar sensing,'' \emph{Proc. IEEE}, vol.~99, no.~7, pp. 1236--1259, 2011.

\bibitem{Ren10153696}
Z.~Ren, L.~Qiu, J.~Xu, and D.~W.~K. Ng, ``Robust transmit beamforming for secure integrated sensing and communication,'' \emph{IEEE Trans. Commun.}, vol.~71, no.~9, pp. 5549--5564, Sep. 2023.

\bibitem{Hua10086626}
H.~Hua, J.~Xu, and T.~X. Han, ``Optimal transmit beamforming for integrated sensing and communication,'' \emph{IEEE Trans. Veh. Technol.}, vol.~72, no.~8, pp. 10\,588--10\,603, 2023.

\bibitem{Liu9724205}
X.~Liu, T.~Huang, and Y.~Liu, ``Transmit design for joint {MIMO} radar and multiuser communications with transmit covariance constraint,'' \emph{IEEE J. Sel. Areas Commun.}, vol.~40, no.~6, pp. 1932--1950, Jun. 2022.

\bibitem{Zhi}
Z.-Q. Luo, W.-K. Ma, A.~M.-C. So, Y.~Ye, and S.~Zhang, ``Semidefinite relaxation of quadratic optimization problems,'' \emph{IEEE Signal Process. Mag.}, vol.~27, no.~3, pp. 20--34, May 2010.

\bibitem{Huang2010new}
Y.~Huang and D.~P. Palomar, ``Rank-constrained separable semidefinite programming with applications to optimal beamforming,'' \emph{IEEE Trans. Signal Process.}, vol.~58, no.~2, pp. 664-- 678, Feb. 2010.

\bibitem{Tuan10214071}
T.~A. Le, D.~W.~K. Ng, and X.-S. Yang, ``A rank-one optimization framework and its applications to transmit beamforming,'' \emph{IEEE Trans. Veh. Technol.}, vol.~73, no.~1, pp. 620--636, Jan. 2024.

\bibitem{Boyd_convex}
S.~Boyd and L.~Vandenberghe, \emph{Convex Optimization}.\hskip 1em plus 0.5em minus 0.4em\relax Cambridge University Press, 2004.

\bibitem{Wei11}
H.~Dahrouj and W.~Yu, ``Multicell interference mitigation with joint beamforming and common message decoding,'' \emph{IEEE Trans. Commun.}, vol.~59, no.~8, pp. 2264--2273, Aug. 2011.

\bibitem{YangFA2008}
X.-S. Yang, \emph{Nature-Inspired Metaheuristic Algorithms}.\hskip 1em plus 0.5em minus 0.4em\relax Luniver Press, 2008.

\bibitem{YangFA2009}
------, \emph{Engineering optimisation: an introduction with metaheuristic applications}.\hskip 1em plus 0.5em minus 0.4em\relax Wiley, 2009.

\bibitem{Tuan10311527}
T.~A. Le and X.-S. Yang, ``Generalized firefly algorithm for optimal transmit beamforming,'' \emph{IEEE Trans. Wireless Commun.}, vol.~23, no.~6, pp. 5863--5877, Jun. 2024.

\bibitem{Tuan10201127}
------, ``Firefly algorithm for beamforming design in {RIS-aided} communication systems,'' in \emph{2023 IEEE 97th Veh. Technol. Conf. (VTC2023-Spring)}, 2023, pp. 1--5.

\bibitem{TuanWCL2024}
M.~K. Hoang, T.~A. Le, K.-X. Thuc, T.~V. Luyen, X.-S. Yang, and D.~W.~K. Ng, ``Firefly algorithm for movable antenna arrays,'' \emph{IEEE Wireless Commun. Letters}, vol.~13, no.~11, pp. 3157--3161, 2024.

\bibitem{Rathapon6331681}
R.~Saruthirathanaworakun, J.~M. Peha, and L.~M. Correia, ``Opportunistic sharing between rotating radar and cellular,'' \emph{IEEE J. Sel. Areas Commun.}, vol.~30, no.~10, pp. 1900--1910, Nov. 2012.

\bibitem{Tuan_VTC2024Spring}
T.~A. Le, I.~Ku, X.-S. Yang, C.~Masouros, and T.~Le-Ngoc, ``Cognitive beamforming design for dual-function radar-communications,'' in \emph{2024 IEEE 99th Veh. Technol. Conf. (VTC2024-Spring)}, 2024, pp. 1--5.

\bibitem{Memisoglu}
E.~Memisoglu, M.~B. Janjua, and H.~Arslan, ``Power-efficient time-domain scheduling for {ISAC} beamforming,'' \emph{IEEE Wireless Commun.etters}, vol.~13, no.~10, pp. 2837--2841, Oct. 2024.

\bibitem{Cao2023}
Y.~Cao and Q.-Y. Yu, ``Joint resource allocation for user-centric cell-free integrated sensing and communication systems,'' \emph{IEEE Commun. Letters}, vol.~27, no.~9, pp. 2338--2342, Sep. 2023.

\bibitem{Abanto-Leon}
L.~F. Abanto-Leon and S.~Maghsudi, ``Optimal user and target scheduling, user-target pairing, and low-resolution phase-only beamforming for {ISAC} systems,'' \emph{IEEE Trans. Veh. Technol.}, pp. 1--6, 2025, early Access.

\bibitem{Stoica}
P.~Stoica, J.~Li, and Y.~Xie, ``On probing signal design for {MIMO} radar,'' \emph{IEEE Trans. Signal Process.}, vol.~55, no.~8, pp. 4151--4161, Aug. 2007.

\bibitem{Yang10605608}
R.~Yang, Z.~Zhu, J.~Zhang, S.~Xu, C.~Li, Y.~Huang, and L.~Yang, ``Deep learning-based joint transmit beamforming for dual-functional radar-communication system,'' \emph{IEEE Trans. Wireless Commun.}, vol.~23, no.~10, pp. 15\,198--15\,211, 2024.

\bibitem{Tuan_CL2018}
H.~Al-Salihi, T.~Van~Chien, T.~A. Le, and M.~R. Nakhai, ``A successive optimization approach to pilot design for multi-cell massive {MIMO} systems,'' \emph{IEEE Commun. Letters}, vol.~22, no.~5, pp. 1086--1089, May 2018.

\bibitem{Tuan_TVT_2020}
T.~A. Le, T.~Van~Chien, M.~R. Nakhai, and T.~Le-Ngoc, ``Pareto-optimal pilot design for cellular massive {MIMO} systems,'' \emph{IEEE Trans. Veh. Technol.}, vol.~69, no.~11, pp. 13\,206--13\,215, Nov. 2020.

\bibitem{Zhu_TDD_2025}
Z.~Zhu, R.~Yang, C.~Li, Y.~Huang, and L.~Yang, ``Adaptive joint sparse {Bayesian} approaches for near-field channel estimation,'' \emph{IEEE Trans. Wireless Commun.}, vol.~24, no.~3, pp. 2590--2605, Mar. 2025.

\bibitem{Yang_TDD_2025}
R.~Yang, S.~Xu, Z.~Zhu, C.~Li, Y.~Huang, and L.~Yang, ``Knowledge-driven channel estimation for asymmetrical massive {MIMO} systems,'' \emph{IEEE Trans. Veh. Technol.}, vol.~74, no.~1, pp. 911--924, Jan. 2025.

\bibitem{Zhang_TDD_2024}
J.~Zhang, S.~Xu, R.~Yang, C.~Li, and L.~Yang, ``{HDnGAN:} a channel estimation method for time-varying {mmWave} massive {MIMO},'' \emph{IEEE Trans. Wireless Commun.}, vol.~23, no.~11, pp. 17\,189--17\,204, Nov. 2024.

\bibitem{ZhaoFDD}
Q.~Zhao, X.~Zeng, Z.~Fan, Q.~Zhang, and W.~Li, ``Channel estimation for {FDD} massive {MIMO} with complex residual denoising network,'' \emph{IEEE Wireless Commun. Letters}, vol.~13, no.~8, pp. 2070--2074, Aug. 2024.

\bibitem{GAOFDD}
Z.~Gao, L.~Dai, W.~Dai, B.~Shim, and Z.~Wang, ``Structured compressive sensing-based spatio-temporal joint channel estimation for {FDD} massive {MIMO},'' \emph{IEEE Trans. Commun.}, vol.~64, no.~2, pp. 601--617, Feb. 2016.

\bibitem{Wesemann_FDD_2017}
S.~Wesemann and T.~L. Marzetta, ``{Channel Training for Analog FDD Repeaters: Optimal Estimators and Cramér–Rao Bounds},'' \emph{IEEE Trans. Signal Process.}, vol.~65, no.~23, pp. 6158--6170, Dec. 2017.

\bibitem{Xiang_FDD_2023}
B.~Xiang, D.~Hu, and J.~Wu, ``Deep learning-based downlink channel estimation for {FDD} massive {MIMO} systems,'' \emph{IEEE Wireless Commun. Letters}, vol.~12, no.~4, pp. 699--702, Apr. 2023.

\bibitem{cvx2015}
{CVX Research Inc.}, ``{CVX}: Matlab software for disciplined convex programming, academic users,'' \url{http://cvxr.com/cvx}, 2015.

\bibitem{TuanTcom2013}
T.~A. Le and M.~R. Nakhai, ``Downlink optimization with interference pricing and statistical {CSI},'' \emph{IEEE Trans. Commun.}, vol.~61, no.~6, pp. 2339--2349, Jun 2013.

\bibitem{Yang_FA_Multi_Modal_2009}
X.-S. Yang, ``Firefly algorithms for multimodal optimization,'' in \emph{Stochastic Algorithms: Foundations and Applications}, O.~Watanabe and T.~Zeugmann, Eds.\hskip 1em plus 0.5em minus 0.4em\relax Berlin, Heidelberg: Springer Berlin Heidelberg, 2009, pp. 169--178.

\bibitem{10475421}
A.~M. Elbir, A.~Abdallah, A.~Celik, and A.~M. Eltawil, ``Antenna selection with beam squint compensation for integrated sensing and communications,'' \emph{IEEE Journal of Selected Topics in Signal Processing}, vol.~18, no.~5, pp. 857--870, Jul. 2024.

\bibitem{10335685}
Y.~Gong, M.~Mahmood, R.~Morawski, and T.~Le-Ngoc, ``Dual-layer metamaterial rectangular antenna arrays for in-band full-duplex massive mimo,'' \emph{IEEE Access}, vol.~11, pp. 135\,708--135\,727, 2023.

\bibitem{10736998}
C.~Yang, W.~Yi, and B.~Champagne, ``Joint antenna selection and beamforming for area surveillance with spatially distributed array radar,'' \emph{IEEE Trans. Aerospace and Electronic Systems}, pp. 1--15, 2024.

\bibitem{Mats}
M.~Bengtsson and B.~Ottersten, ``Optimal downlink beamforming using {S}emidefinite optimization,'' in \emph{Proc. 37th Annu. Allerton Conf. Commun., Control, and Comput.}, 1999, pp. 987 -- 996.

\bibitem{Tuanglobecom11}
T.~A. Le and M.~R. Nakhai, ``An iterative algorithm for downlink multi-cell beam-forming,'' in \emph{Proc. IEEE Global Telecommun. Conf.}, Dec. 2011, pp. 1--6.

\bibitem{Yongwei}
Y.~Huang and D.~P. Palomar, ``Rank-constrained separable {S}emidefinite programming with applications to optimal beamforming,'' \emph{IEEE Trans. Signal Process.}, vol.~58, no.~2, pp. 644--678, Feb. 2010.

\end{thebibliography}

\begin{IEEEbiography}[{\includegraphics[width=1.in,height=2.8in,clip,keepaspectratio]{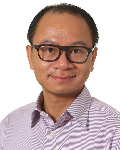}}]{Tuan Anh Le (Senior, IEEE)} received the Ph.D. degree in telecommunications research from King’s College London, The University of London, U.K., in 2012. His Ph.D. research contributed to the Green Radio project funded by the Core 5 joint research program of the U.K.’s 
Engineering and Physical Sciences Research Council (EPSRC) and the Virtual Center of Excellence in Mobile and Personal Communications (Mobile VCE). 

He was a Post-Doctoral Research Fellow with the School of Electronic and Electrical Engineering, University of Leeds, Leeds, U.K. He is a Senior Lecturer at Middlesex University, London, U.K.  His current research interests include integrated sensing and communication (ISAC), Fluid antenna systems (FAS), Reconfigurable-intelligent-surface (RIS)-aided communications, Radio frequency (RF) energy harvesting and wireless power transfer, physical-layer security, nature-inspired optimization, and applied machine learning for wireless communications. He was a Technical Program Chair of the 26th International Conference on Telecommunications (ICT 2019). He was an Exemplary Reviewer of IEEE Communications Letters in 2019 and 2023. Dr. Le serves as an Editor of IEEE Wireless Communications Letters. He was honored as the Best Editor of 2024 of IEEE Wireless Communications Letters.
\end{IEEEbiography}
\begin{IEEEbiography}[{\includegraphics[width=1in,height=1.25in]{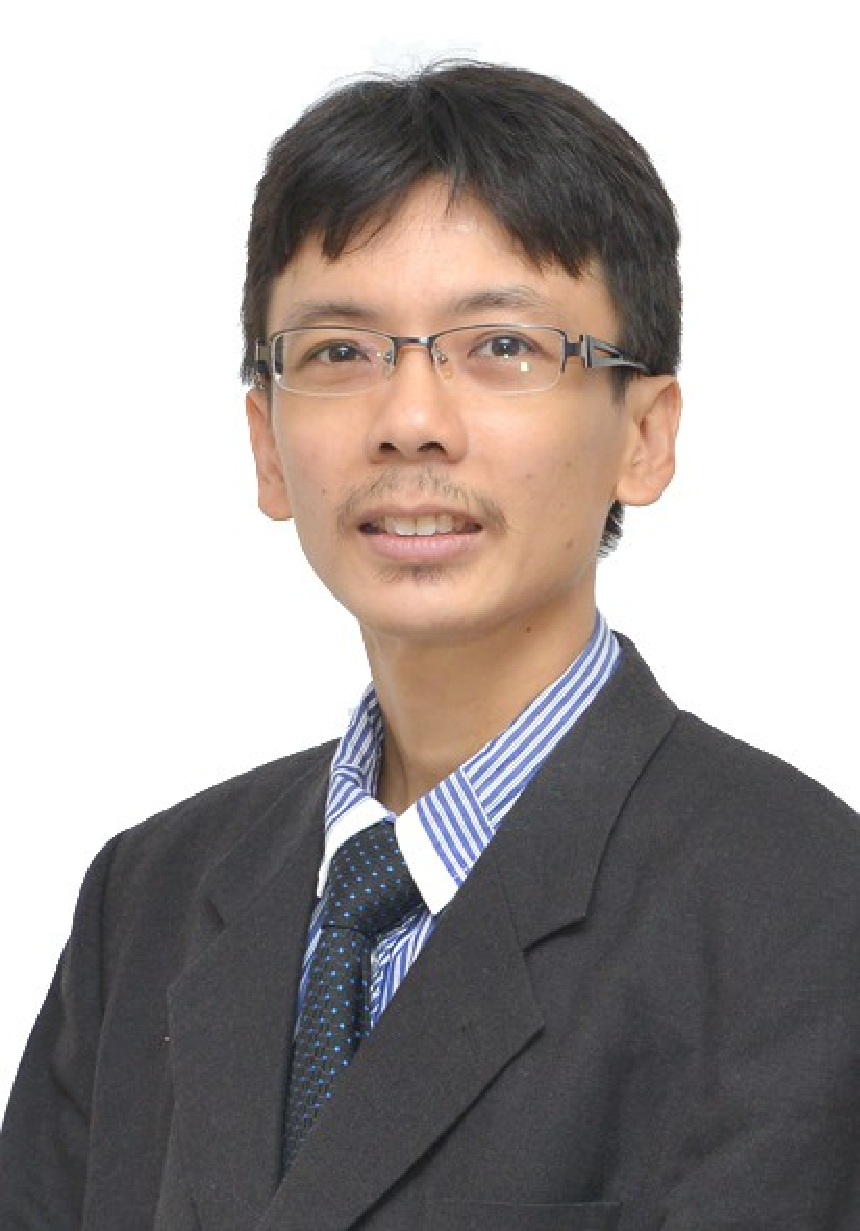}}]{Ivan Ku}received his B.Eng. degree in Electronics and M.Eng.Sc. degree in Telecommunications from Multimedia University, Cyberjaya, Malaysia, in 2001 and 2006, respectively. He earned his Ph.D. in Wireless Communications in 2013 through a joint research program between Heriot-Watt University and the University of Edinburgh, United Kingdom. His doctoral studies were supported by the Mobile VCE (Virtual Center of Excellence), a collaborative U.K. research initiative in mobile communications funded by industry and government agencies. In 2017, he was awarded the Erasmus Mundus LEADERS scholarship (Staff Mobility Level), funded by the European Union under the Erasmus Mundus program. He is also a member of the 5G IMT Sub-Working Group under the Malaysian Technical Standards Forum. Currently, he serves as an Assistant Professor in the Faculty of Artificial Intelligence and Engineering at Multimedia University. His research interests include cooperative MIMO systems, joint sensing and communication (JSAC), cognitive radio networks, and green communications.
\end{IEEEbiography}
\begin{IEEEbiography}[{\includegraphics[width=1.in,height=2.8in,clip,keepaspectratio]{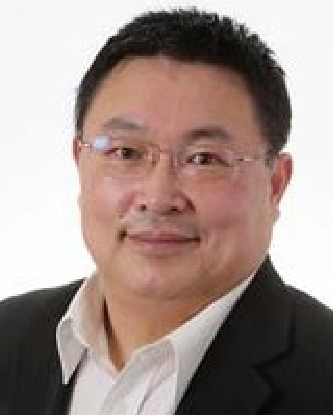}}]{Xin-She Yang} is Reader in Modelling and Simulation at Middlesex University London. After receiving his DPhil in Applied Mathematics from University of Oxford in 1998, he then worked at Cambridge University and later as a Senior Research Scientist at National Physical Laboratory (UK). He is also an elected Fellow of the Institute of Mathematics and its Applications, Fellow of Asian Computational Intelligence Society and Fellow of Soft Computing Research Society. He is the co-Editor of Springer’s book series: Springer Tracts in Nature-Inspired Computing and has been on editorial boards of multiple international journals. He is also a member of the UKRI Talent Peer Review College. He was the IEEE Computational Intelligence Society (CIS) chair for the Task Force on Business Intelligence and Knowledge Management (2015 to 2020). He has authored/edited more than 50 books and has published more than 400 peer-reviewed research papers with more than 98,000 citations. He has been on the prestigious list of most influential researchers or highly-cited researchers (Web of Science) for ten consecutive years since 2016.
\end{IEEEbiography}
\begin{IEEEbiography}[{\includegraphics[width=1.in,height=2.8in,clip,keepaspectratio]{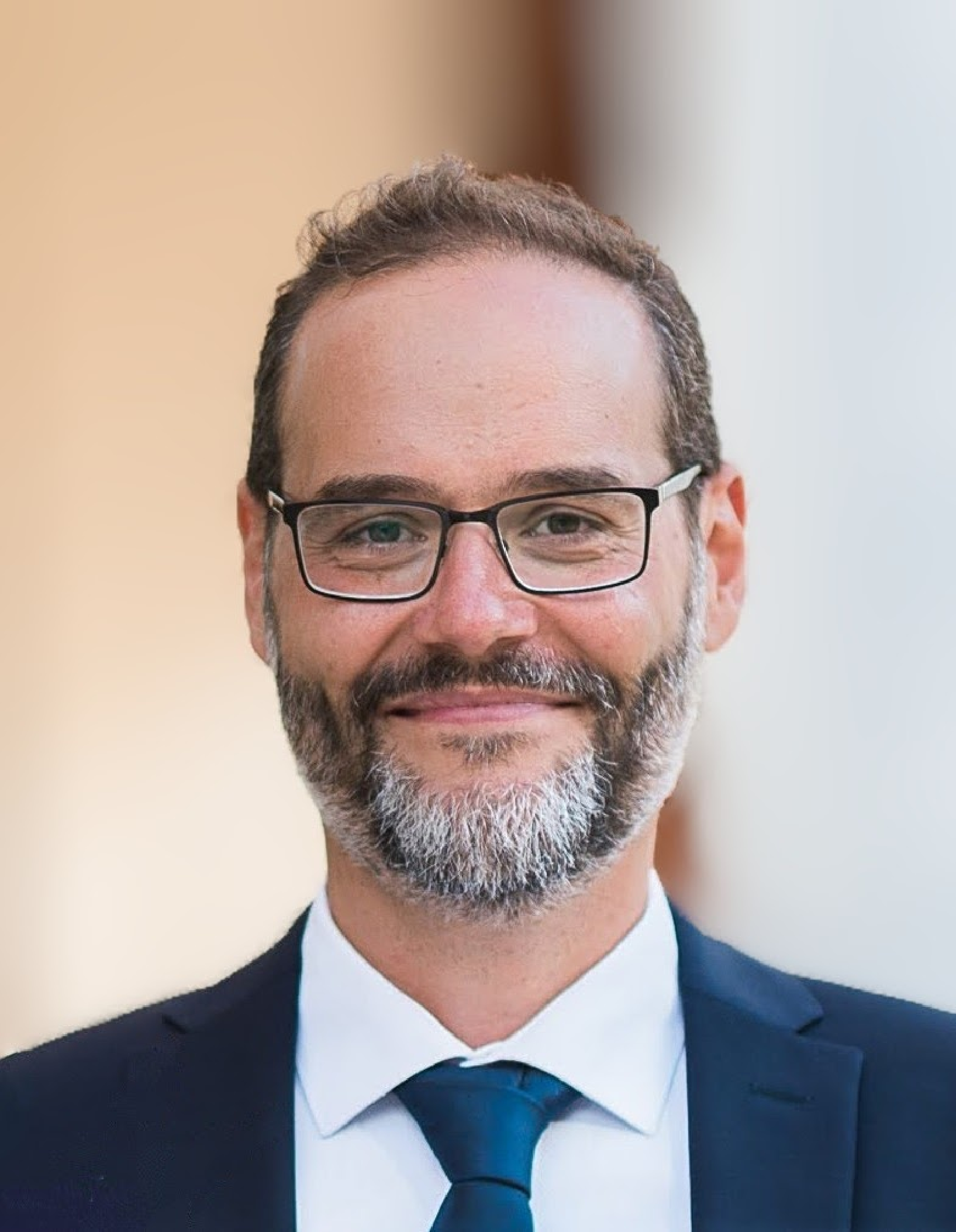}}]{Christos Masouros (Fellow, IEEE, Fellow, IET)} received the Diploma degree in Electrical and Computer Engineering from the University of Patras, Greece, in 2004, and MSc by research and PhD in Electrical and Electronic Engineering from the University of Manchester, UK in 2006 and 2009 respectively. In 2008 he was a research intern at Philips Research Labs, UK, working on the LTE standards. Between 2009-2010 he was a Research Associate in the University of Manchester and between 2010-2012 a Research Fellow in Queen's University Belfast. In 2012 he joined University College London as a Lecturer. He has held a Royal Academy of Engineering Research Fellowship between 2011-2016. 

Since 2019 he is a Full Professor of Signal Processing and Wireless Communications in the Information and Communication Engineering research group, Dept. Electrical and Electronic Engineering, and affiliated with the Institute for Communications and Connected Systems, University College London. His research interests lie in the field of wireless communications and signal processing with particular focus on Green Communications, Large Scale Antenna Systems, Integrated Sensing and Communications, interference mitigation techniques for MIMO and multicarrier communications. Between 2018-22 he was the Project Coordinator of the €4.2m  EU H2020 ITN project PAINLESS, involving 12 EU partner universities and industries, towards energy-autonomous networks. Between 2024-28 he will be the Scientific Coordinator of the €2.7m EU H2020 DN project ISLANDS, involving 19 EU partner universities and industries, towards next generation vehicular networks. He is a Fellow of the IEEE, Fellow of the Insitute of Electronic Engineers (IET), the Artificial Intelligence Industry Alliance (AIIA) and the Asia-Pacific Artificial Intelligence Association (AAIA). He was the recipient of the 2024 IEEE SPS Best Paper Award, the 2024 IEEE SPS Donald G. Fink Overview Paper Award, the 2023 IEEE ComSoc Stephen O. Rice Prize, co-recipient of the 2021 IEEE SPS Young Author Best Paper Award and the recipient of the Best Paper Awards in the IEEE GlobeCom 2015 and IEEE WCNC 2019 conferences.  He is an IEEE ComSoc Distinguished lecturer 2024-2025,  and his work on ISAC has been featured in the World Economic Forum’s report on the top 10 emerging technologies. He has been recognised as an Exemplary Editor for the IEEE Communications Letters, and as an Exemplary Reviewer for the IEEE Transactions on Communications. He is an Area Editor for IEEE Transactions on Wireless Communications, and Editor-at-Large for IEEE Open Journal of the Communications Society. He has been an Editor for IEEE Transactions on Communications, IEEE Transactions on Wireless Communications, the IEEE Open Journal of Signal Processing, Associate Editor for IEEE Communications Letters, and a Guest Editor for a number of IEEE Journal on Selected Topics in Signal Processing issues. He is a founding member and Vice-Chair of the IEEE Emerging Technology Initiative on Integrated Sensing and Communications (SAC), Chair of the IEEE SPS ISAC Technical Working Group, and Chair of the IEEE Green Communications \& Computing Technical Committee, Special Interest Group on Green ISAC.  He is a member of the IEEE Standards Association Working Group on ISAC performance metrics, and a founding member of the ETSI ISG on ISAC. He was the TPC chair for the IEEE ICC 2024 Selected Areas in Communications (SAC) Track on ISAC, Chair of the IEEE PIMRC2024 Track 1 on PHY and Fundamentals, Chair of the "Integrated Imaging and Communications" stream in IEEE CISA 2024, and TPC Co-Chair of IEEE VTC 2025.
\end{IEEEbiography}
\begin{IEEEbiography}[{\includegraphics[width=1in,height=1.25in]{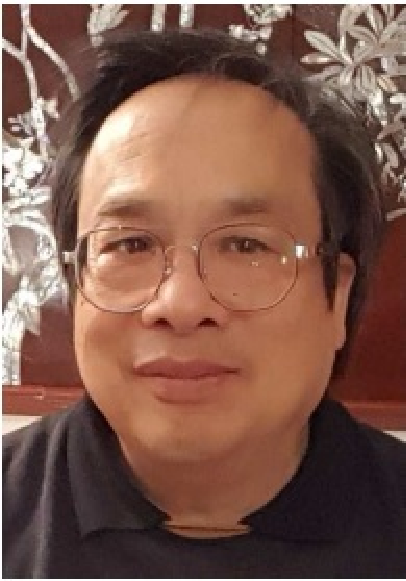}}]{Tho Le-Ngoc (Life Fellow, IEEE)} received the B.Eng. degree in electrical engineering and the M.Eng. degree in microprocessor applications from McGill University, Montreal, QC, Canada, in 1976 and 1978, respectively, and the Ph.D. degree in digital communications from the University of Ottawa, Ottawa, ON, Canada, in 1983. 

From 1977 to 1982, he was with Spar Aerospace Ltd., Sainte-Anne-de-Bellevue, QC, Canada, involved in the development and design of satellite communications systems. From 1982 to 1985, he was with SRTelecom, Inc., Saint Laurent, QC, Canada, where he developed the new point-to-multipoint DA-TDMA/TDM Subscriber Radio System SR500. From 1985 to 2000, he was a Professor with the Department of Electrical and Computer Engineering, Concordia University, Montreal.

Since 2000, he has been with the Department of Electrical and Computer Engineering, McGill University. His research interest includes broadband digital communications. Dr. Le-Ngoc was a recipient of the 2004 Canadian Award in Telecommunications Research and the IEEE Canada Fessenden Award in 2005. He is a Distinguished James McGill Professor, and a Fellow of the Engineering Institute of Canada, the Canadian Academy of Engineering, and the Royal Society of Canada.
\end{IEEEbiography}
\end{document}